\documentclass[useAMS,amssymb,usenatbib]{mn2e}
\usepackage{epsfig}
\usepackage{color}
\usepackage{graphicx}
\usepackage{tabularx}
\usepackage{longtable}
\usepackage[draft]{hyperref}           
\usepackage[normalem]{ulem}

\usepackage{xcolor}
\usepackage[percent]{overpic}

\def\tol#1#2#3{\hbox{\rule{0pt}{15pt}${#1}^{+{#2}}_{-{#3}}$}}

\newcommand{\Ms}{\mathrm{M}_\odot}

\newcommand{\LCDM}{\Lambda\mathrm{CDM}}
\newcommand{\vm}{\mathrm{v_{max}}}
\newcommand{\rmax}{\mathrm{r_{max}}}
\newcommand{\rh}{\mathrm{r_{1/2}}}
\newcommand{\vh}{\mathrm{v_{1/2}}}
\newcommand{\vc}{\mathrm{v_{c}}}

\newcommand{\eagle}{{\sc Eagle }}

\title[The APOSTLE simulations]{The APOSTLE simulations: solutions to the Local Group's cosmic puzzles}

\author[Sawala et al.]  {\parbox{\textwidth}{Till
    Sawala$^{1,2}$\thanks{E-mail: \texttt{till.sawala@durham.ac.uk}},
    Carlos~S.~Frenk$^1$,  Azadeh Fattahi$^3$, Julio~F.~Navarro$^{3,4}$,
    Richard~G.~Bower$^1$, Robert~A.~Crain$^5$, Claudio Dalla
    Vecchia$^{6,7}$, Michelle Furlong$^1$, John.~C.~Helly$^1$, Adrian
    Jenkins$^1$, Kyle~A.~Oman$^2$, Matthieu Schaller$^1$, Joop
    Schaye$^8$, Tom Theuns$^{1}$, James Trayford$^1$ and Simon~D.~M.~White$^9$}\vspace{0.4cm}\\
  \parbox{\textwidth}{ $^{1}$Institute for Computational Cosmology,
    Department of Physics,
    University of Durham, South Road, Durham DH1 3LE, UK \\
    $^{2}$Department of Physics, University of Helsinki, Gustaf
    H\"allstr\"omin katu 2a,
    FI-00014 Helsinki, Finland\\
    $^{3}$Department of Physics and Astronomy, University of Victoria,
    3800 Finnerty Road, Victoria, British Columbia V8P 5C2, Canada\\
    $^{4}$Senior CIfAR Fellow\\
    $^{5}$Astrophysics Research Institute, Liverpool John Moores
    University, 146 Brownlow Hill, Liverpool L3 5RF, UK\\
    $^{6}$Instituto de Astrof\'{i}sica de Canarias, C/ V\'{i}a  L\'{a}ctea s/n, 38205 La Laguna, Tenerife, Spain \\
    $^{7}$Departamento de Astrof\'isica, Universidad de La Laguna, Av.~del Astrof\'isico Francisco S\'anchez s/n, 38206 La Laguna, Tenerife, Spain
    $^{8}$Leiden Observatory, Leiden University, Postbus  9513, 2300 RA Leiden, The Netherlands \\
    $^{9}$Max-Planck Institute for Astrophysics,
    Karl-Schwarzschild-Straße 1, 85741 Garching, Germany }}

\begin{document}

\date{Accepted 2015 ***. Received 2015 ***; in original
  form 2015}

\pagerange{\pageref{firstpage}--\pageref{lastpage}} \pubyear{2015}

\maketitle

\label{firstpage}

\begin{abstract}
  The Local Group of galaxies offer some of the most discriminating
  tests of models of cosmic structure formation. For example,
  observations of the Milky Way (MW) and Andromeda satellite
  populations appear to be in disagreement with N-body simulations of
  the ``Lambda Cold Dark Matter'' ($\Lambda$CDM) model: there are far
  fewer satellite galaxies than substructures in cold dark matter
  halos (the ``missing satellites'' problem); dwarf galaxies seem to
  avoid the most massive substructures (the ``too-big-to-fail''
  problem); and the brightest satellites appear to orbit their host
  galaxies on a thin plane (the ``planes of satellites''
  problem). Here we present results from {\sc Apostle} (A Project Of
  Simulating The Local Environment), a suite of cosmological
  hydrodynamic simulations of twelve volumes selected to match the
  kinematics of the Local Group (LG) members. Applying the \eagle code
  to the LG environment, we find that our simulations match the
  observed abundance of LG galaxies, including the satellite galaxies
  of the MW and Andromeda. Due to changes to the structure of halos
  and the evolution in the LG environment, the simulations reproduce
  the observed relation between stellar mass and velocity dispersion
  of individual dwarf spheroidal galaxies without necessitating the
  formation of cores in their dark matter profiles. Satellite systems
  form with a range of spatial anisotropies, including one similar to
  that of the MW, confirming that such a configuration is not
  unexpected in $\Lambda$CDM. Finally, based on the observed velocity
  dispersion, size, and stellar mass, we provide new estimates of the
  maximum circular velocity for the halos of nine MW dwarf
  spheroidals.
\end{abstract}


\section{Introduction} \label{sec:Introduction} The ability of the
cold dark matter (CDM) model to predict observables on different
scales and at different epochs lies at the root of its remarkable
success. Predictions for the anisotropy of the microwave background
radiation \citep{Peebles-1987}, and the large scale distribution of
galaxies \citep{Davis-1985}, were made soon after the model was
formulated, and have since been spectacularly validated by
observations \citep{Planck-2013}. However, observations on scales
currently testable only within the Local Group (LG) have yielded
results that appear to be in conflict with CDM predictions.

The ``missing satellites'' problem \citep{Klypin-1999,Moore-1999}
refers to the apparent paucity of luminous satellite galaxies compared
to the large number of dark matter substructures predicted in
$\Lambda$CDM\footnote{Throughout this paper, we use $\Lambda$CDM to
  refer to the $\Lambda$ Cold Dark Matter model, where $\Lambda$
  refers to the cosmological constant model for dark energy, which is
  constrained primarily by observations outside the Local Group. We
  also assume that the dark matter is collisionless.} . That the
number of observed dwarf galaxies does not directly mirror the number
of substructures is perhaps no surprise: it has long been predicted
that processes such as supernova feedback \citep[e.g.][]{Larson-1974},
and cosmic reionisation \citep[e.g.][]{Efstathiou-1992} should reduce
the star formation efficiency in low mass halos, and even prevent the
smallest ones from forming stars altogether.

The potential of these processes to bring the stellar mass function in
$\Lambda$CDM into agreement with observations has already been
demonstrated using semi-analytical models \citep[e.g.][]{Benson-2002,
  Somerville-2002, Font-2011, Guo-2011}. However, these do not
investigate whether the resulting satellites would also match the
observed kinematics, or indeed predict ratios between stellar mass and
maximum circular velocity, $\vm$, that are far lower than
observed. Extending the success of $\Lambda$CDM from the overall
galaxy population down to the number of observed Local Group dwarf
galaxies through direct simulations has remained a challenge.

Furthermore, the apparent excess of substructures in $\Lambda$CDM is
not just limited to the lowest mass scales: simulations also predict
the presence of subhalos so massive that they should not be affected
by reionisation (and hence deemed ``too big to fail'',
\citep{Boylan-Kolchin-2011}), but whose internal structure seems
incompatible with that of the brightest observed satellites. While the
two main galaxies within the LG both have several satellite galaxies
whose rotation curves are consistent with massive subhalos, their
number falls short of the number of such substructures predicted to
surround Milky Way (MW) or M31 mass halos in $\Lambda$CDM.

Finally, it has long been known that most of the eleven brightest, so
called ``classical'' MW satellites, appear to orbit the Galaxy on a
thin plane, and a similar (but distinct) plane has subsequently also
been identified among satellites of Andromeda. While $\Lambda$CDM
satellite systems are known to be anisotropic, the satellite systems
of the MW and Andromeda are purported to be extremely unusual
\citep{Pawlowski-2013}.

The problems, at times reported as fatal for $\Lambda$CDM, arise when
observations are confronted with predictions from dark matter only
(DMO) simulations that treat the cosmic matter content as a single
collisionless fluid, a poor approximation on scales where baryonic
processes are important. It has, of course, long been recognised that
the distribution of light is not a precise tracer of dark matter, but
simple models for populating dark matter structures with galaxies do
not capture the complexity of galaxy formation physics. On the other
hand, hydrodynamic simulations have confirmed the importance of
baryonic effects, but have either focussed on individual dwarf
galaxies, ignoring the LG environment \citep[e.g.][]{Stinson-2009,
  Governato-2010, Sawala-2010, Sawala-2011, Nickerson-2011, Shen-2013,
  Wheeler-2015}, or have not yet reproduced the observed LG galaxy
population \citep[e.g.][]{Benitez-Llambay-2014}. The lack of a single
model able to reconcile all of the LG observations with $\Lambda$CDM
predictions has led to renewed interest in alternatives to CDM, such
as warm \citep[e.g.][]{Lovell-2012} or self-interacting dark matter
\citep[e.g.][]{Rocha-2013, Vogelsberger-2014a}.

Here we test the $\Lambda$CDM model with a new suite of cosmological
hydrodynamic simulations, with initial conditions tailored to match
the LG environment. In particular, we focus on pairs of halos that
match the separation, approach velocity, and relative tangential
velocity of the Milky Way (MW) and Andromeda (M31). From a large
cosmological simulation, we have selected twelve pairs of halos with
combined virial masses of $\sim2.3\pm0.6\times 10^{12} \Ms$,
compatible with the most recent dynamical constraints
\citep{Gonzalez-2013, Penarrubia-2014}. The selection and set-up of
our initial conditions are described in more detail by
\cite{Fattahi-2015}.

We have re-simulated each LG volume at several resolutions, both as
dark matter only (DMO) simulations, and as hydrodynamic simulations,
with the code developed for the {\it Evolution and Assembly of
  GaLaxies and their Environments} project ({\sc Eagle},
\cite{Schaye-2014, Crain-2015}).

In this paper, we give an introduction to the simulations, and compare
our results to the observed LG galaxy population. We show that the
abundance of galaxies in the LG can be reproduced within $\Lambda$CDM
using a galaxy formation model calibrated on much larger scales. By
comparing the DMO and hydrodynamic simulations, we show that the
apparent discrepancies between observations and $\Lambda$CDM
predictions can be resolved once baryonic effects are included.

This paper is organised as follows. In Section~\ref{sec:Methods}, we
give a brief overview of the \eagle code
(Section~\ref{sec:Methods-Code}), followed by a description of the
{\sc Apostle} volumes (Section~\ref{sec:Methods-ICs}). In
Section~\ref{sec:Results}, we present our key results, and
implications for the questions outlined above: the stellar mass
function (Section~\ref{sec:Results-SMF}), and the satellites ``too big
to fail'' (Section~\ref{sec:Results-TBTF}). We demonstrate that our
simulations also produce the right satellite galaxies in the right
satellite halos (Section~\ref{sec:right}), and provide estimates for
$\vm$ of observed satellites based on the measured velocity dispersion
and stellar mass (Section~\ref{sec:vmax}). We revisit the apparent
kinematical and spatial alignment of satellites in
Section~\ref{sec:Results-Plane} and conclude with a summary in
Section~\ref{sec:Summary}.

\section{Methods}\label{sec:Methods}
\subsection{The EAGLE Galaxy Formation Model} \label{sec:Methods-Code}
The simulations presented in this paper were performed with the code
developed for the {\it Evolution and Assembly of GaLaxies and their
  Environments} ({\sc Eagle}, \cite{Schaye-2014, Crain-2015})
project. The \eagle code is a substantially modified version of {\sc
  P-Gadget-3}, which itself is an improved version of the publicly
available {\sc Gadget-2} code \citep{Springel-2005}. Gravitational
accelerations are computed using the standard Tree-PM scheme of {\sc
  P-Gadget-3}, while hydrodynamic forces are computed in the smoothed
particle hydrodynamics (SPH) scheme of {\sc Anarchy} described in
Dalla-Vecchia et al. (in prep.) and \cite{Schaller-2015b} , which uses
the pressure-entropy formalism introduced by \cite{Hopkins-2013}.

\eagle is an evolution of the models used in the {\sc Gimic}
\citep{Crain-2009} and {\sc Owls} \citep{Schaye-2010} projects and has
been calibrated to reproduce the $z=0.1$ stellar mass function and
galaxy sizes from $10^8\Ms$ to $10^{11}\Ms$ in a cosmological volume
of $100^3$ Mpc$^3$. In addition, the \eagle simulations also
successfully reproduce many other properties and scaling laws of
observed galaxy populations, including the evolution of the stellar
mass function \citep{Furlong-2015}, and the luminosities and colours
of galaxies \citep{Trayford-2015}.

The subgrid physics model of \eagle is described in detail by
\cite{Schaye-2014}. It includes radiative cooling, star formation,
stellar evolution and stellar mass loss, and thermal feedback that
captures the collective effects of stellar winds, radiation pressure
and supernova explosions. It also includes black hole growth fuelled
by gas accretion and mergers, and feedback from active galactic nuclei
(AGN) \citep{Booth-2009, Rosas-Guevara-2013}. Within the Local Group,
AGN feedback is negligible, and the main processes regulating the
formation of galaxies are gas cooling and heating by the UV
background, star formation, and supernova feedback, which are
described in more detail below.

Following \cite{Wiersma-2009}, net cooling rates are computed
separately for 11 elements, assuming ionisation equilibrium in the
presence of uniform UV and X-ray backgrounds from quasars and galaxies
\citep{Haardt-2001}, and the cosmic microwave background
(CMB). Hydrogen is assumed to reionise instantaneously at $z=11.5$,
which is implemented by turning on the ionising background. At higher
redshifts the background is truncated at 1 Ryd, limiting its effect to
preventing the formation of molecular hydrogen. During reionization an
extra 2 eV per proton mass are injected to account for the increase in
the photoheating rates of optically thick gas over the optically thin
rates that are used otherwise. For hydrogen this is done at $z=11.5$,
ensuring that the gas is quickly heated to $10^4$~K, but for HeII the
extra heat is distributed in time following a Gaussian centered at
$z=3.5$ with $\sigma(z) = 0.5$, which reproduces the observed thermal
history \citep{Schaye-2000, Wiersma-2009b}. In order to prevent
artificial fragmentation of the ISM, a temperature floor is imposed on
the gas through a polytropic equation of state with index $\gamma =
4/3$, normalised to $T = 8 \times10^3$K at a gas threshold density of
$n_\mathrm{H} = 10^{-1}$cm$^{-3}$ \citep{Schaye-2008}.

The star formation rate is assumed to be pressure-dependent
\citep{Schaye-2008} and follows the observed Kennicutt-Schmidt star
formation law with a metallicity-dependent density threshold
\citep{Schaye-2004}. Energy feedback from star formation is
implemented using the stochastic, thermal prescription of
\cite{DallaVecchia-2012}. The expectation value for the energy
injected per unit stellar mass formed decreases with the metallicity
of the gas and increases with the gas density to account for
unresolved radiative losses and to help prevent spurious numerical
losses. The injected energy is calibrated to reproduce the observed,
$z=0$ galaxy stellar mass function and sizes \citep{Crain-2015}. On
average it is close to the energy available from core collapse
supernovae alone \citep{Schaye-2014}. Galactic winds develop
naturally, without imposing mass loading factors, velocities or
directions.

In our highest resolution simulations, each of the main galaxies
contains more than $20$ million particles, comparable to the best
simulations of individual MW sized galaxies published to
date. Nevertheless, they still barely resolve the scale height of the
MW thin disk, and the effective resolution is also limited by the
equation of state imposed on the gas. Furthermore, the resolution and
the physics included in the \eagle code do not resolve individual
star forming regions or supernova feedback events. We rely instead on
a well-calibrated subgrid physics model to parametrise the star
formation and feedback processes.  For our study, we have used the
same parameter values that were used in the $100^3$Mpc$^3$ L100N1504
\eagle reference simulation \citep{Schaye-2014} independently of
resolution.

While there is clearly scope for future improvements, the relevant
properties discussed in this paper, such as the stellar mass function
and the circular velocity function of substructures are well
converged, as demonstrated in the Appendix. This indicates that our
numerical resolution is sufficient to capture the physical mechanisms
of structure formation, gas accretion and outflows in the \eagle
model.

\subsection{The APOSTLE simulations} \label{sec:Methods-ICs} Our
twelve Local Group regions are zoom simulations selected from a DMO
simulation of $100^3$Mpc$^3$ with $1620^3$ particles in WMAP-7
cosmology \citep{Komatsu-2011}. The resimulation volumes were selected
to match the available dynamical constraints of the Local Group. Each
volume contains a pair of halos in the virial mass\footnote{For halo
  masses, we generally quote $m_{200}$, the mass enclosed in a
  spherical volume whose mean overdensity is 200$\times$ the critical
  density.} range $5\times10^{11} \Ms$ to $2.5\times10^{12} \Ms$, with
median values of $1.4 \times10^{12}\Ms$ for the primary (more massive)
halo and $0.9 \times10^{12}\Ms$ for the secondary (less massive) halo
of each pair. The combined masses of the primary and secondary range
from $1.6\times10^{12}\Ms$ to $3.6\times10^{12}\Ms$ with a median mass
of $2.3\times10^{12}\Ms$, consistent with recent estimates of
$2.40_{-1.05}^{+1.95}\times10^{12}\Ms$ from dynamical arguments and
CDM simulations \citep{Gonzalez-2013}, or $2.3 \pm 0.7 \times 10^{12}$
based on equations of motions that take into account the observed
velocities of galaxies in the local volume \citep{Penarrubia-2014}.

\begin{figure*}
\begin{center}
\begin{overpic}[height=4.35cm]{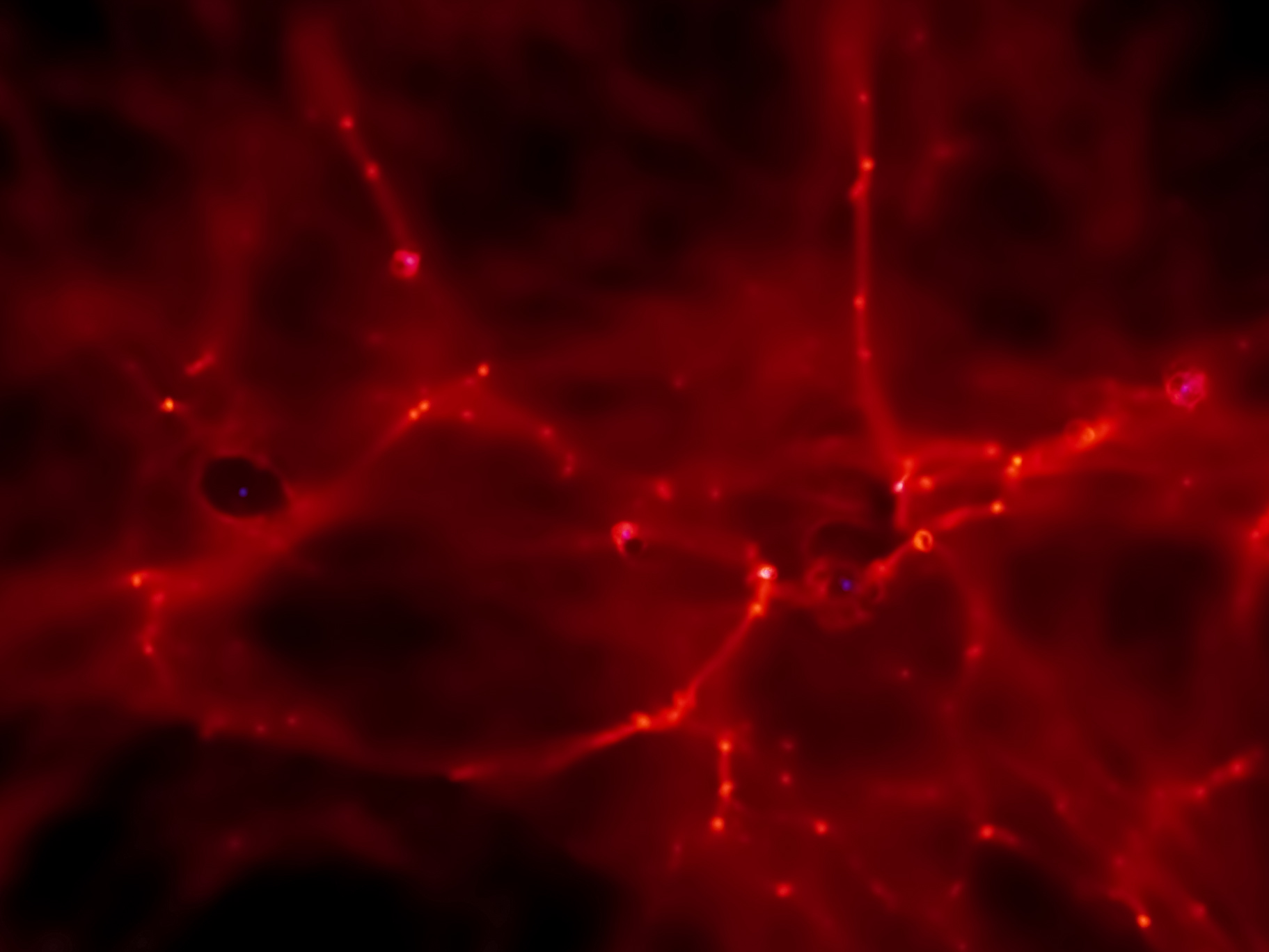}  \put(4,68){\color{white}z=12} \put(3,3){\color{white}\rule{0.223077cm}{1.5pt}} \end{overpic}
\begin{overpic}[height=4.35cm]{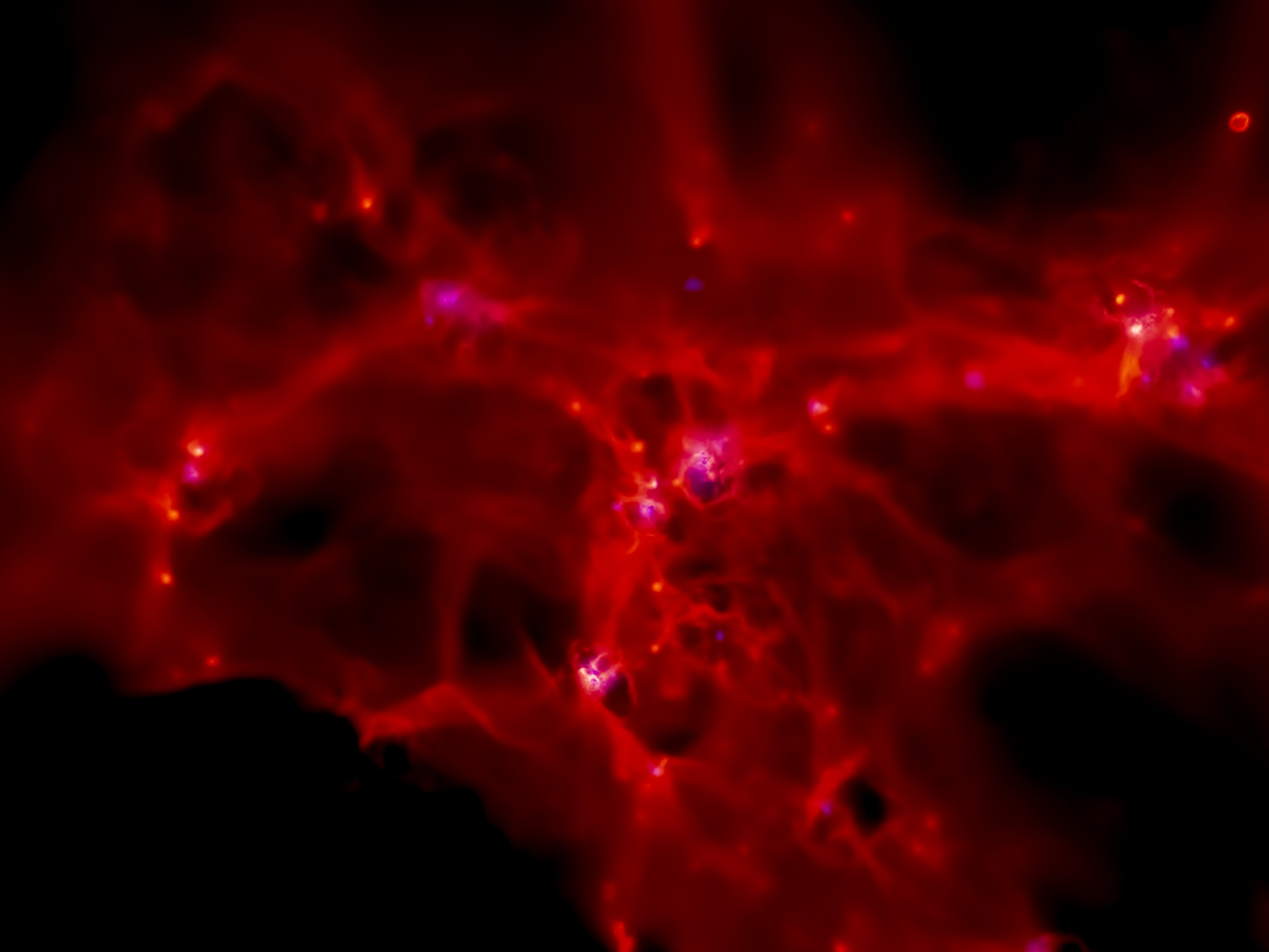}  \put(4,68){\color{white}z=9}  \put(3,3){\color{white}\rule{0.29 cm}{1.5pt}} \end{overpic}
\begin{overpic}[height=4.35cm]{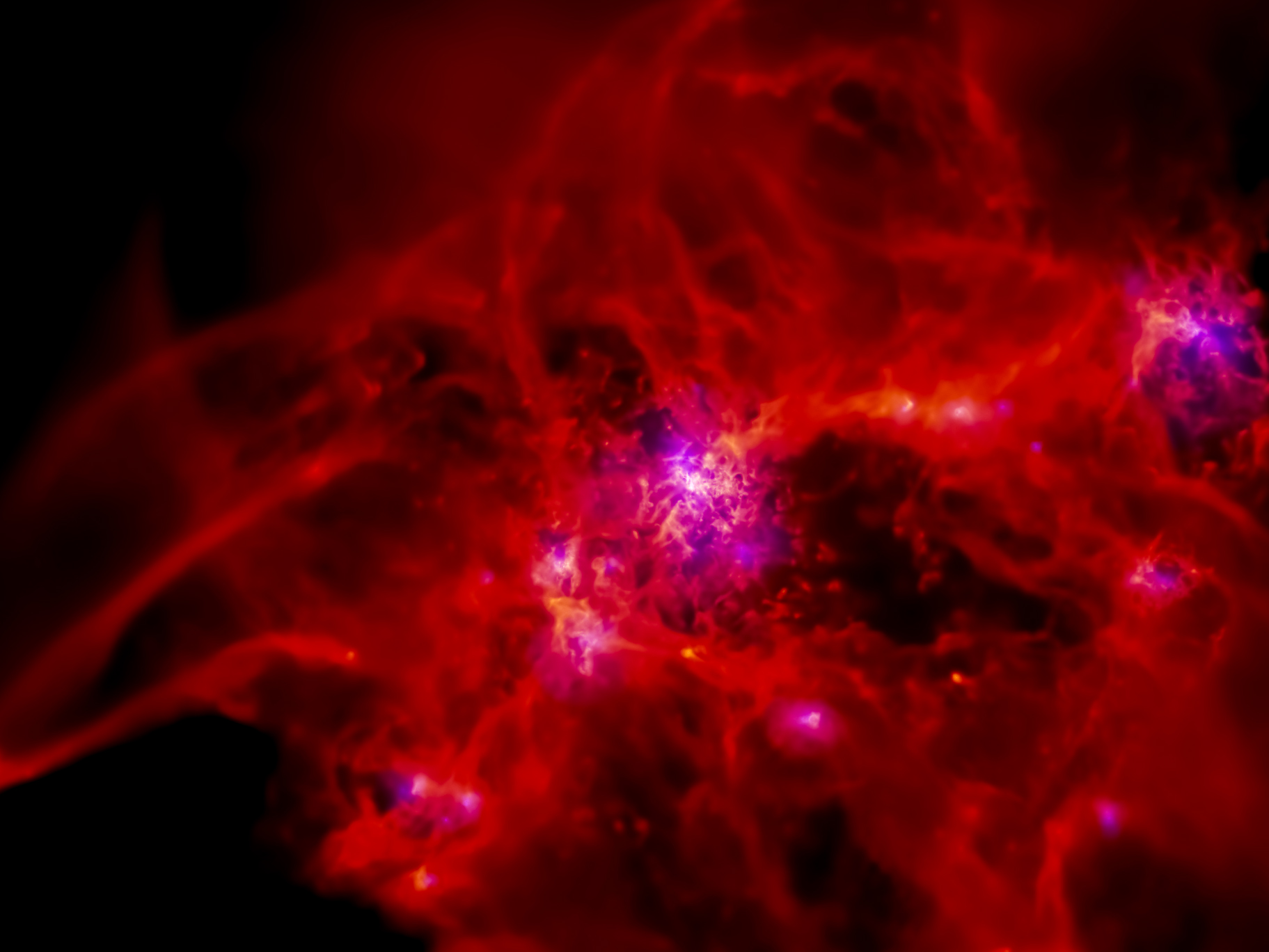}  \put(4,68){\color{white}z=6}  \put(3,3){\color{white}\rule{0.4142857 cm}{1.5pt}} \end{overpic} \\
\addvspace{0.05cm}
\begin{overpic}[height=4.35cm]{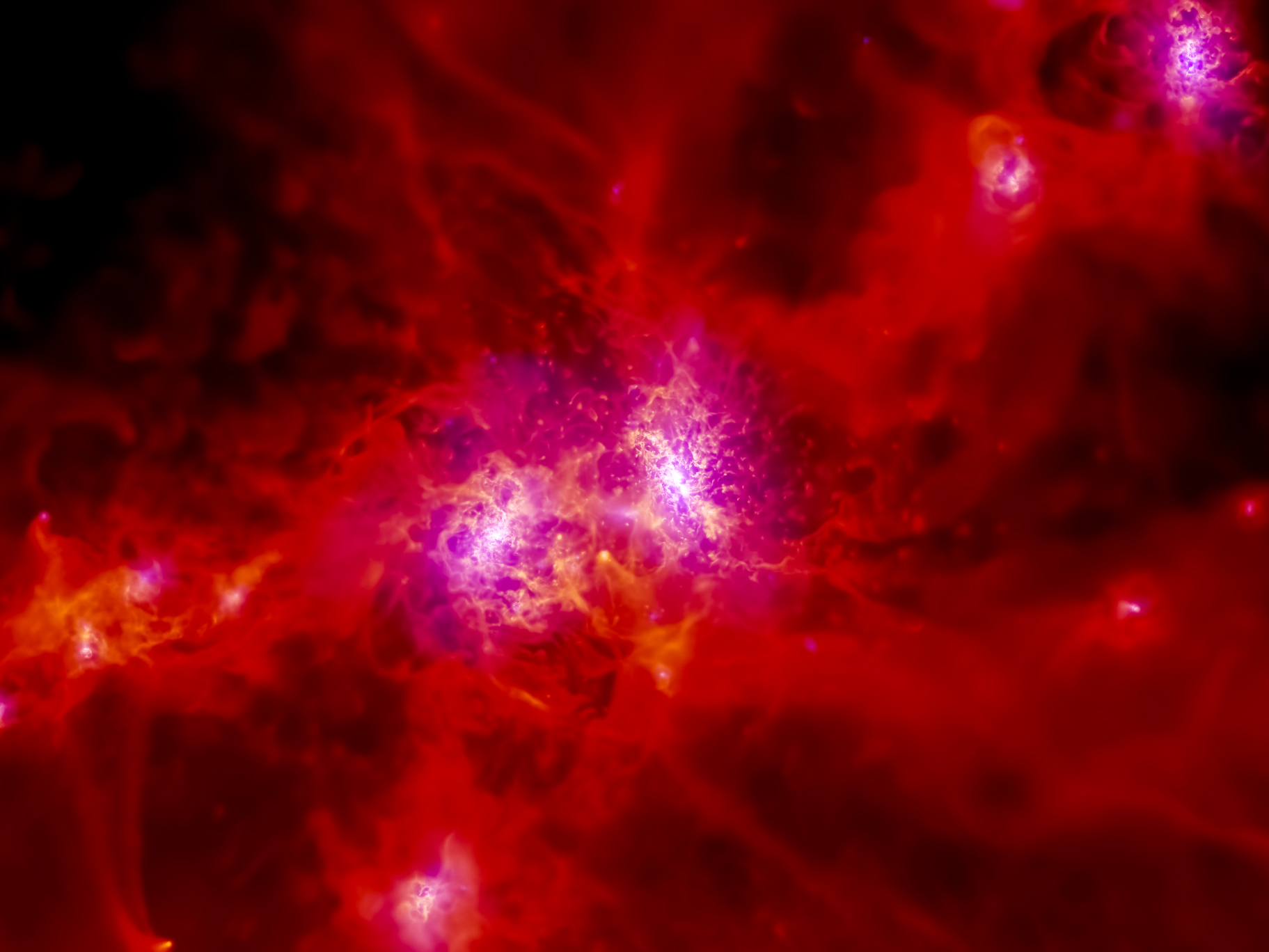}  \put(4,68){\color{white}z=3}  \put(3,3){\color{white}\rule{0.725 cm}{1.5pt}} \end{overpic}
\begin{overpic}[height=4.35cm]{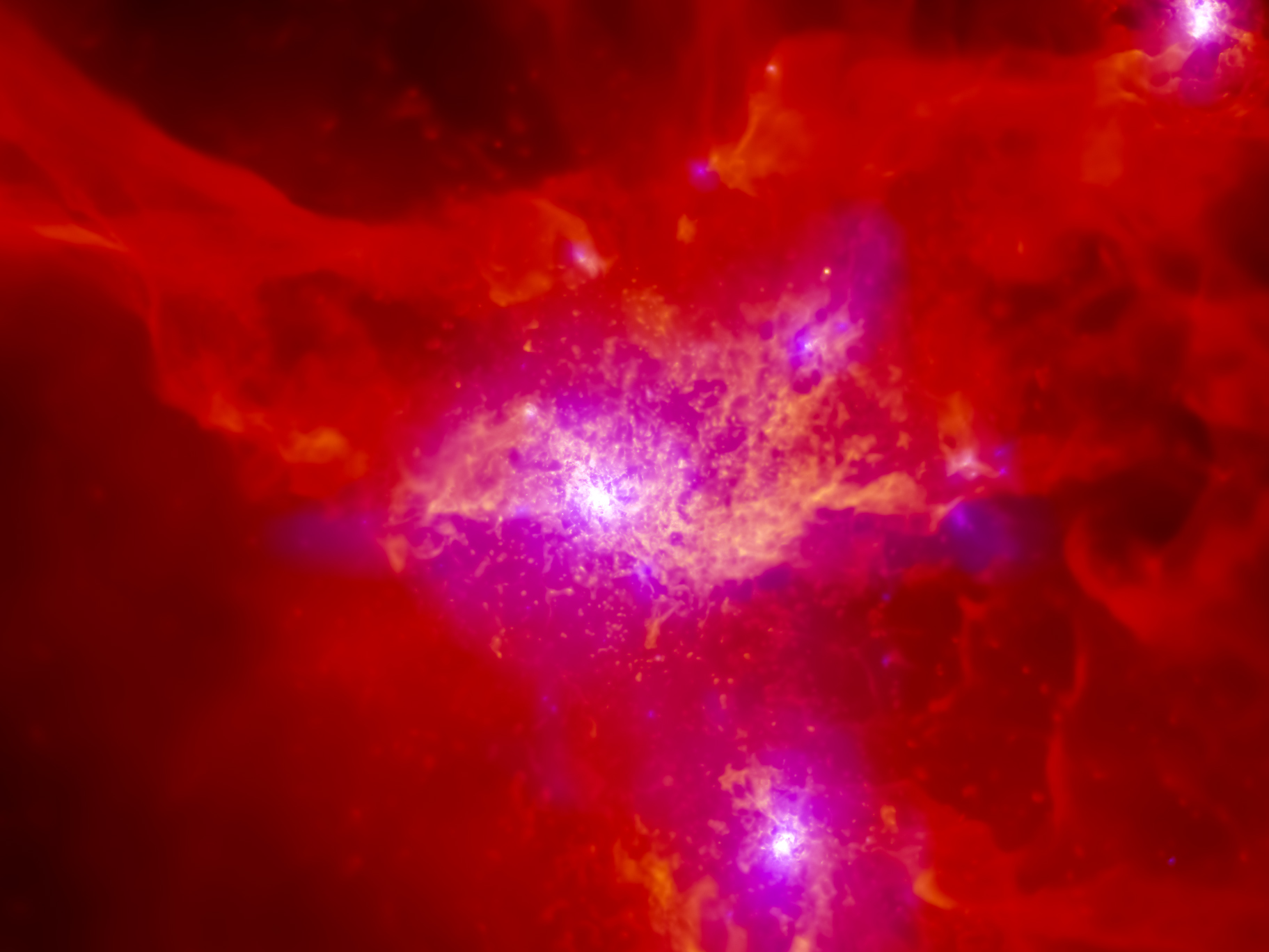}  \put(4,68){\color{white}z=2}  \put(3,3){\color{white}\rule{0.96666 cm}{1.5pt}} \end{overpic}
\begin{overpic}[height=4.35cm]{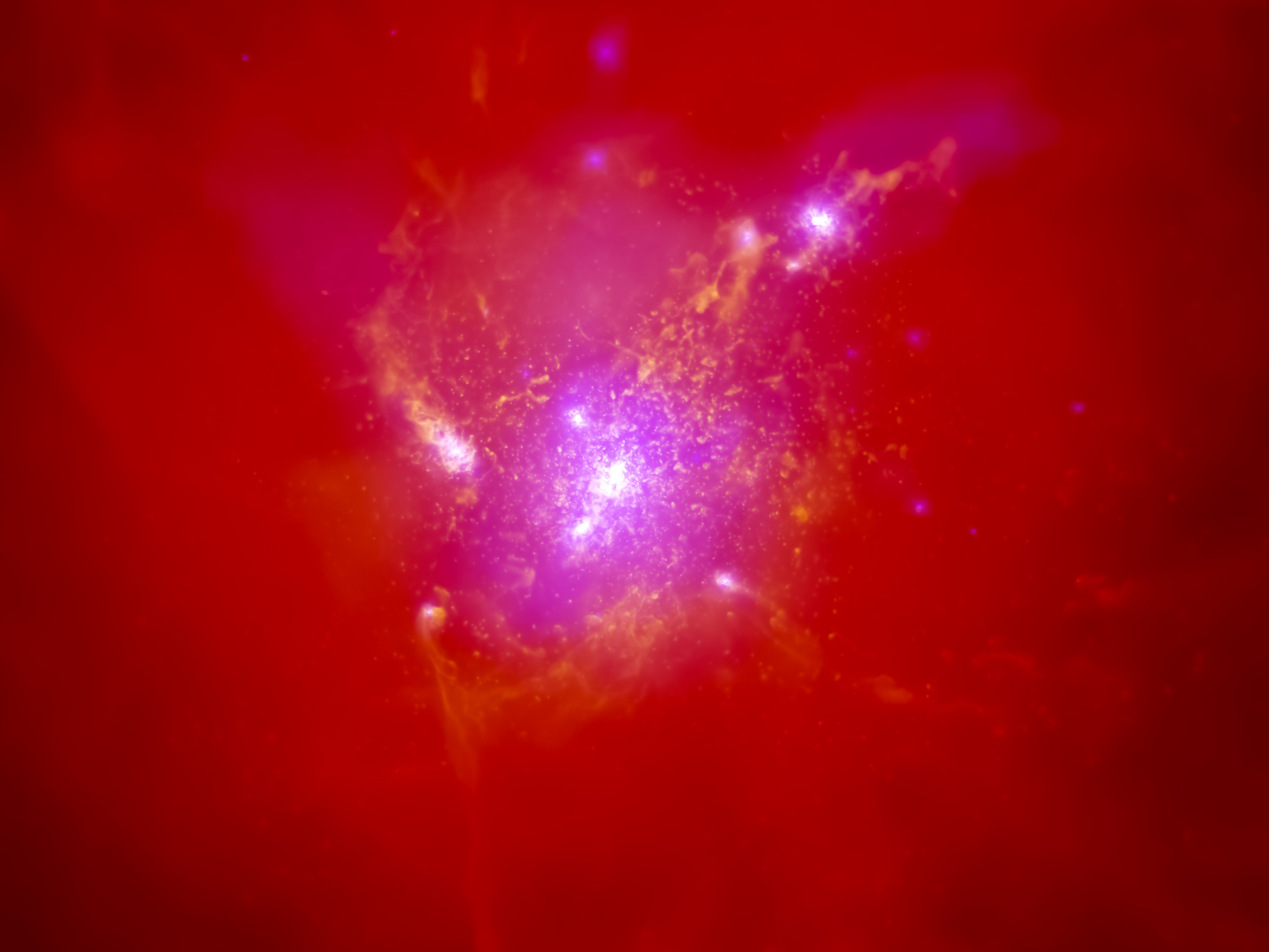}  \put(4,68){\color{white}z=1} \put(3,3){\color{white}\rule{1.45 cm}{1.5pt}}  \end{overpic} \\
\addvspace{0.05cm}
\begin{overpic}[height=4.35cm]{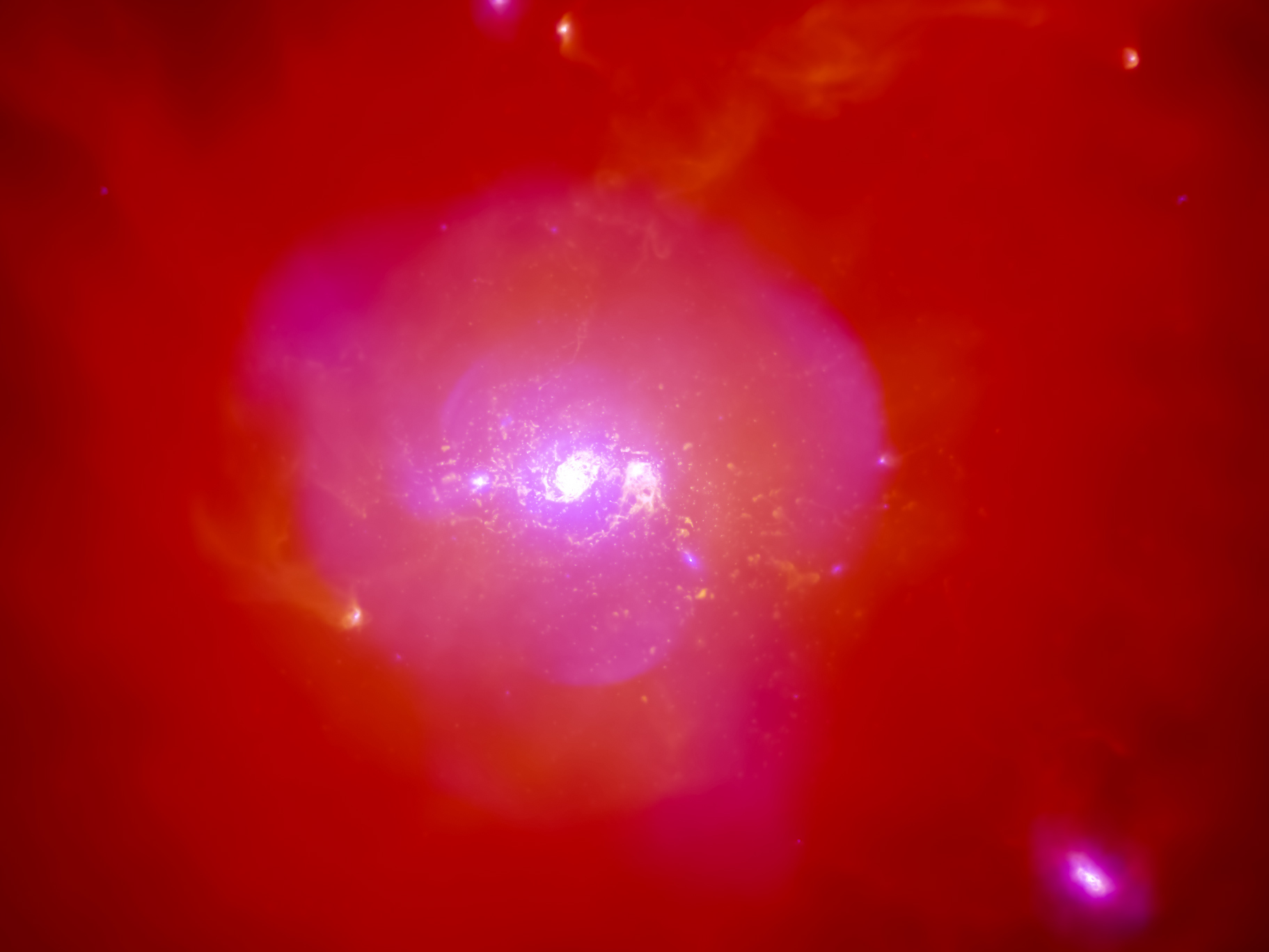}
  \put(4,68){\color{white}z=0.5}   \put(3,3){\color{white}\rule{1.933 cm}{1.5pt}} \end{overpic}
\begin{overpic}[height=4.35cm]{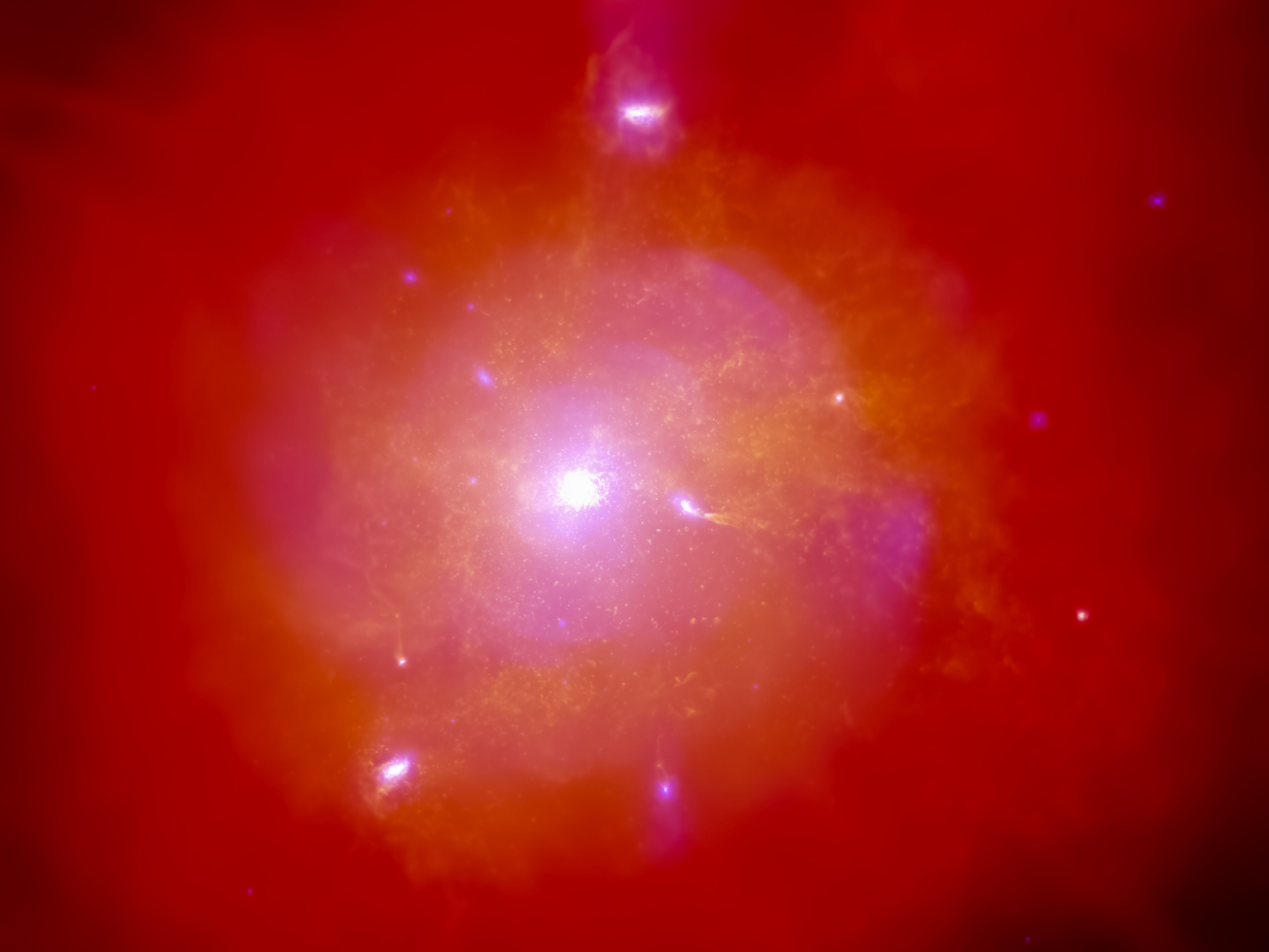}  \put(4,68){\color{white}z=0.25}  \put(3,3){\color{white}\rule{2.32 cm}{1.5pt}} \end{overpic}
\begin{overpic}[height=4.35cm]{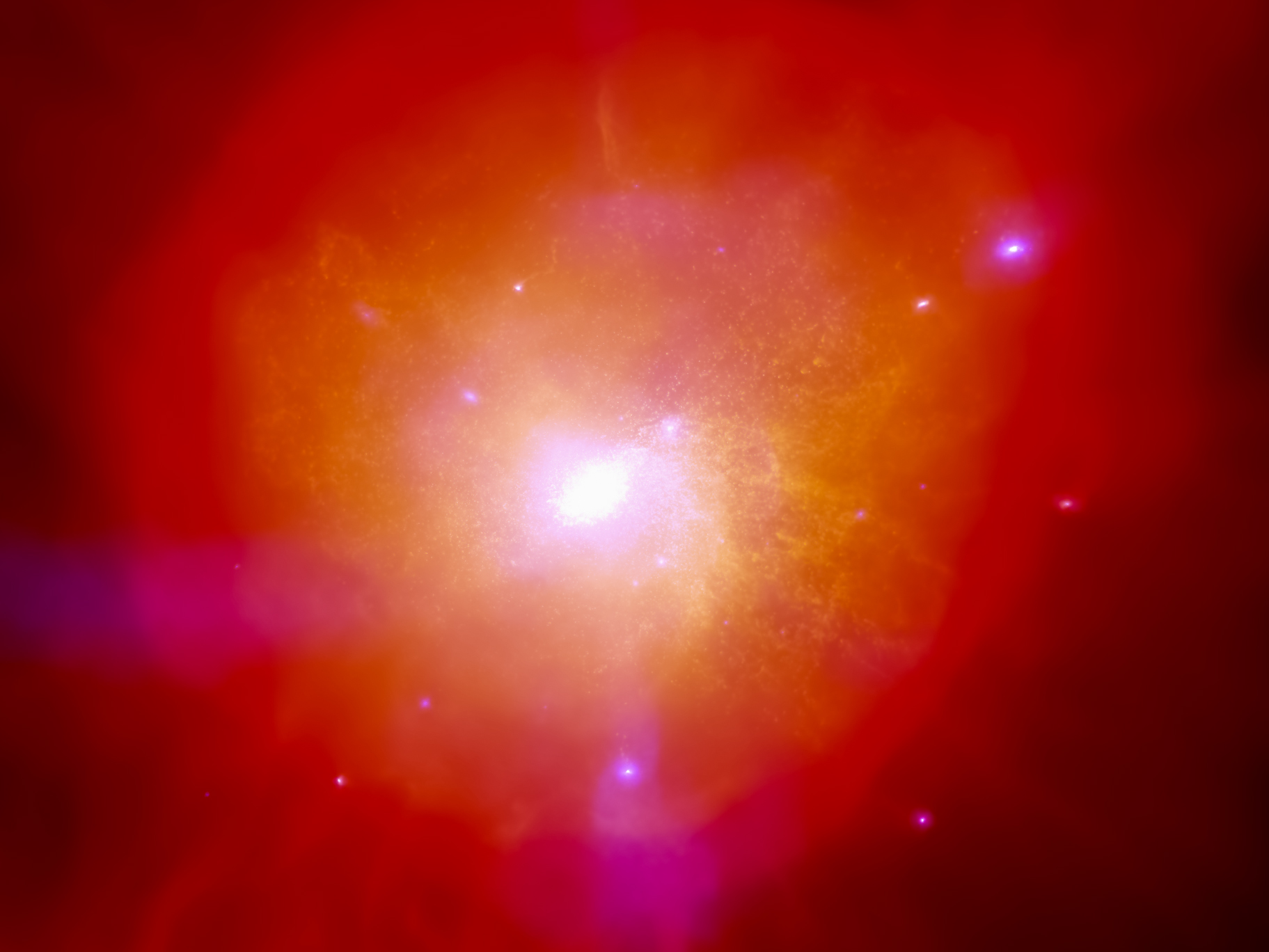}  \put(4,68){\color{white}z=0} \put(3,3){\color{white}\rule{2.9cm}{1.5pt}} \end{overpic} \\
\addvspace{0.2cm}
\begin{overpic}[height=4.35cm, trim=35 0 35 0, clip]{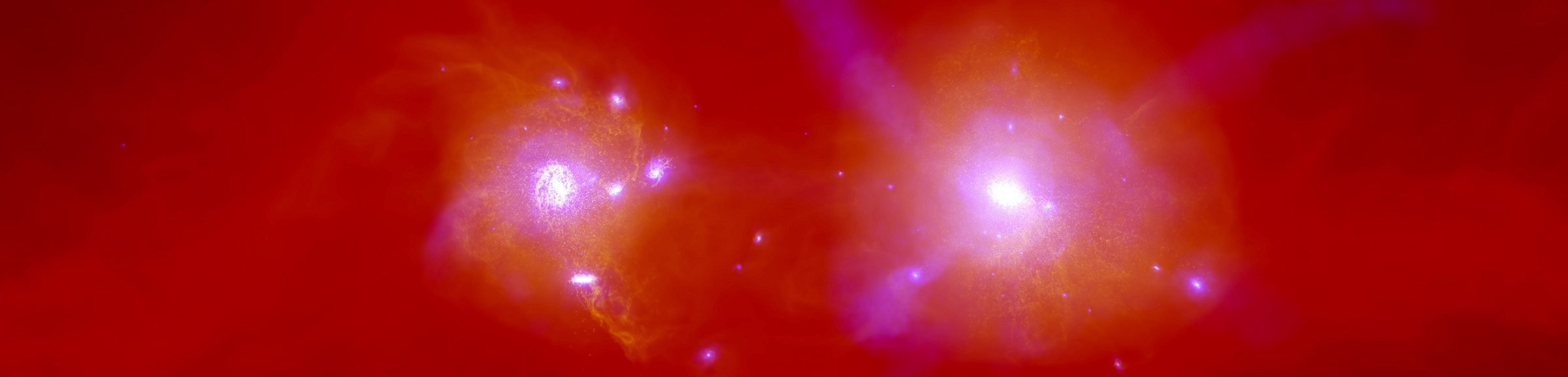}  \put(1.5,1.5){\color{white}z=0} \end{overpic} 

\caption{Top three rows: evolution of gas density (red colours) and
  stellar density (blue colours) in a comoving volume of side length
  $150 \times 200$ h$^{-1}$ kpc, centred on one of the main LG
  galaxies and its progenitors at resolution L1. For scale, a bar of
  length $100 h^{-1}$ kpc is shown on each panel. At $z=12$, before
  reionisation, stars have already formed in some of the highest
  density regions, and feedback from supernovae has begun to blow
  bubbles into the gas. Over time stars form in many more halos, and
  star forming regions merge to form larger galaxies. By $z=1$, the
  main galaxy has formed, and continues to accrete both gas and
  satellites, many of which lose their gas on infall and are also
  tidally disrupted. At $z=0$, the central galaxy is surrounded by
  many satellite galaxies, and a complex stellar halo with visible
  shells and streams. Bottom row: slice through the entire LG in the
  same simulation.\label{fig:evolution} }
\end{center}
\end{figure*}

\begin{figure*}
\begin{center}

\begin{overpic}[width=17.7cm]{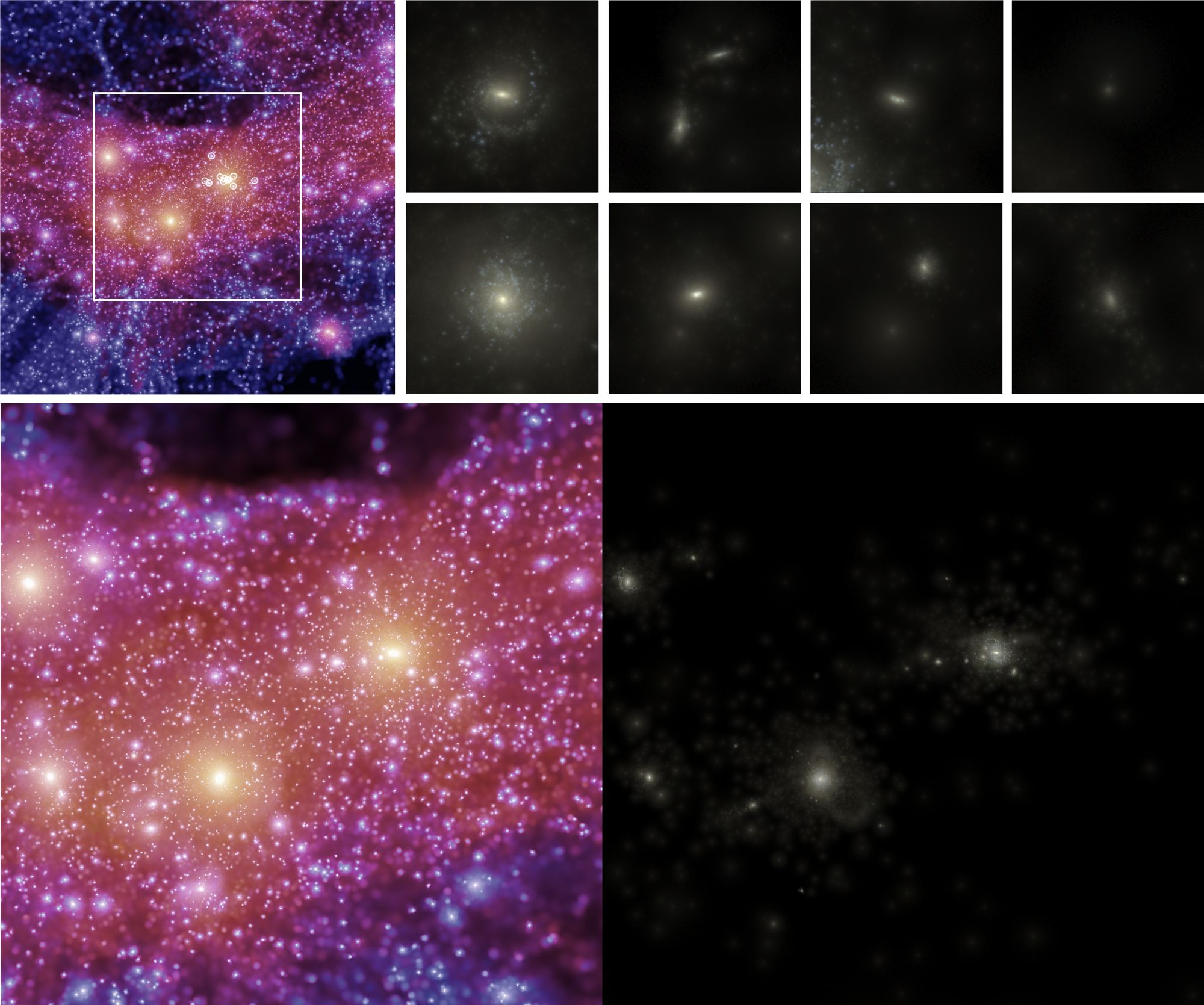} 
\put(3,3){\color{white}Dark matter} 
\put(53,3){\color{white}Galaxies} 
\end{overpic} 
\caption{Top left: projected dark matter density in one of our
  resimulations at resolution level L2 in a cube of side length
  4~Mpc. Circles indicate the locations of the eleven brightest
  satellites of one of the main galaxies, whose spatial distribution
  is as anisotropic as that of the eleven brightest Milky Way
  satellites, and which align with the filament that contains most of
  the halos and galaxies in the region. The main panels contrast the
  vast number of dark matter substructures (left) with the stellar
  light distribution (right) in the 2~Mpc cube indicated by the square
  in the top left panel. The small panels in the top row are of side
  length 125 kpc and reveal in more detail the stellar component of
  some of the different types of galaxies formed in this simulation;
  central galaxies (first and second columns) and satellite galaxies
  (third and fourth columns), which have realistic sizes, colours, and
  morphologies. Dark matter substructures are abundant in the {\sc
    Apostle} simulations, but due to the complexity of galaxy
  formation, starlight paints a very different picture.
} \label{FigLGImage}
\end{center}
\end{figure*}


We further require that the two halos be separated by $800 \pm 200$
kpc, approaching with radial velocity of $\left(0-250\right)$ km/s and
with tangential velocity below $100$ km/s; to have no additional halo
larger than the smaller of the pair within 2.5 Mpc from the midpoint
of the pair, and to be in environments with a relatively unperturbed
Hubble flow out to 4 Mpc. More details about the selection criteria,
and implications of the different dynamical constraints on the total
mass of the LG may be found in \cite{Fattahi-2015}.

The high resolution initial conditions were created using second-order
Lagrangian perturbation theory \citep{Jenkins-2010}. The cosmological
parameters and the linear phases of the parent volume, which are based
on the public multi-scale Gaussian white noise field {\sc Panphasia},
are given in Tables 1 and 6 of \cite{Jenkins-2013}, who also describes
the method used to make the Local Group zoom initial conditions.

Each region sampled with baryons and at the highest resolution
comprises a sphere of at least 2.5~Mpc radius from the LG barycentre
at $z=0$.  Outside of these regions, dark matter particles of
increasing mass are used to sample the large scale environment of the
$100^3$~Mpc$^3$ parent simulation. To investigate the impact of
baryons, we also repeated all our simulations as dark matter only
(DMO), where the dark matter particle masses in the high resolution
region are larger by a factor of
$(\Omega_\mathrm{b}+\Omega_{\mathrm{DM}})/\Omega_{\mathrm{DM}}$ than
in the corresponding hydrodynamic simulations.

The three different resolution levels of the {\sc Apostle} simulations
labelled ``L1'', ``L2'' and ``L3'' have primordial gas (DM) particle
masses of approximately $1.0 (5.0) \times 10^4 \Ms$, $1.2 (5.9)
\times10^5 \Ms $ and $1.5 (7.5) \times10^6 \Ms$, respectively, and
maximum gravitational softening lengths of 134 pc, 307 pc and 711
pc. L3 is close to the resolution of the \eagle L100N1504
simulation. While the \eagle simulations use the Planck-1 cosmology
\citep{Planck-2013}, {\sc Apostle} was performed in the slightly
different WMAP-7 cosmology \citep{Komatsu-2011}, with density
parameters at $z=0$ for matter, baryons and dark energy of
$\Omega_\mathrm{M}=0.272$, $\Omega_\mathrm{b}=0.0455$ and
$\Omega_\mathrm{\lambda}=0.728$, respectively, a Hubble parameter of
H$_0=70.4$~km/s Mpc$^{-1}$, a power spectrum of (linear) amplitude on
the scale of 8$h^{-1}$Mpc of $\sigma_8=0.81$ and a power-law spectral
index $n_s=0.967$. On LG scales, we expect the effect of cosmological
parameters to be minimal.

\begin{figure*}
\begin{center}
\includegraphics [height=6.6cm, trim=2.5cm .8cm 1.8cm 1.5cm, clip]{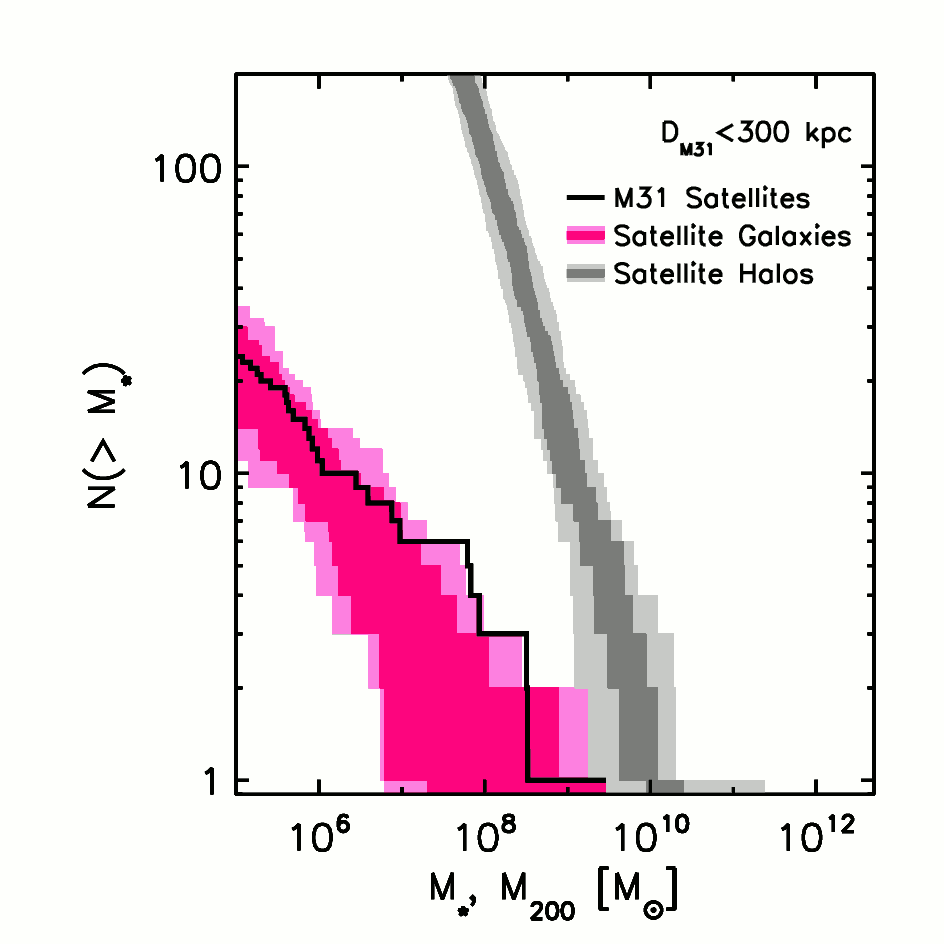} 
\includegraphics [height=6.6cm, trim=5.5cm .8cm 1.8cm 1.5cm, clip]{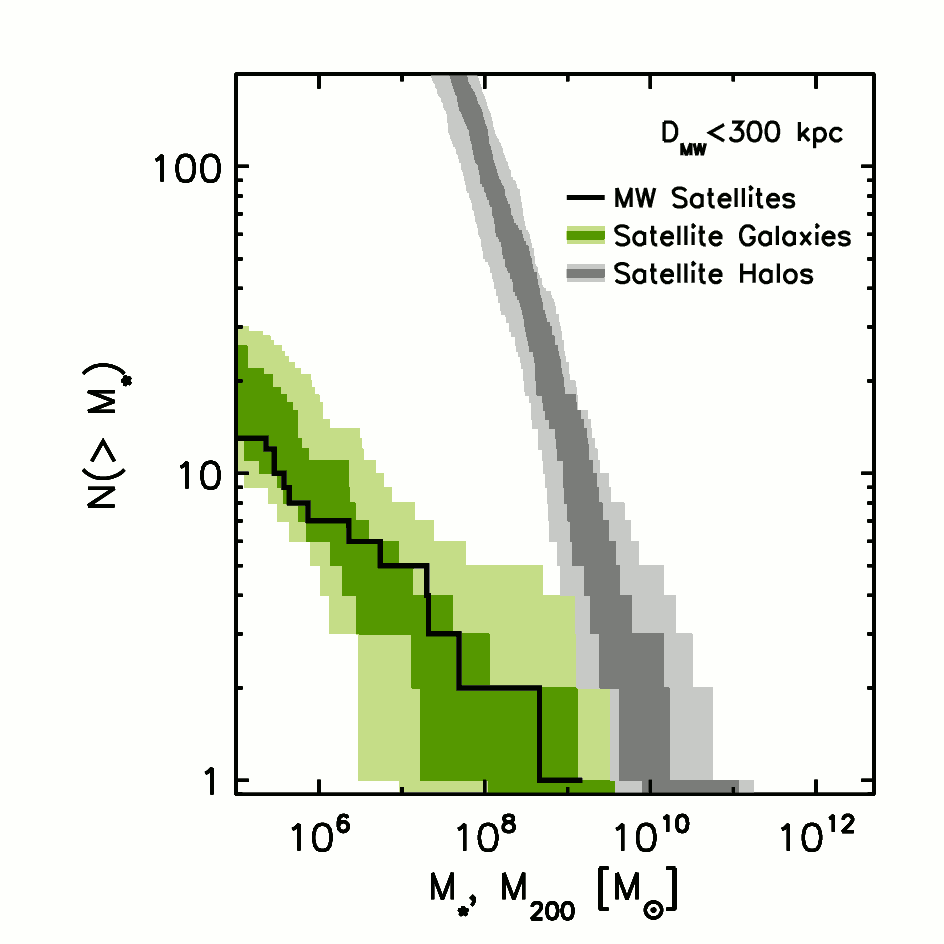}
\includegraphics [height=6.6cm, trim=5.5cm .8cm 1.8cm 1.5cm, clip]{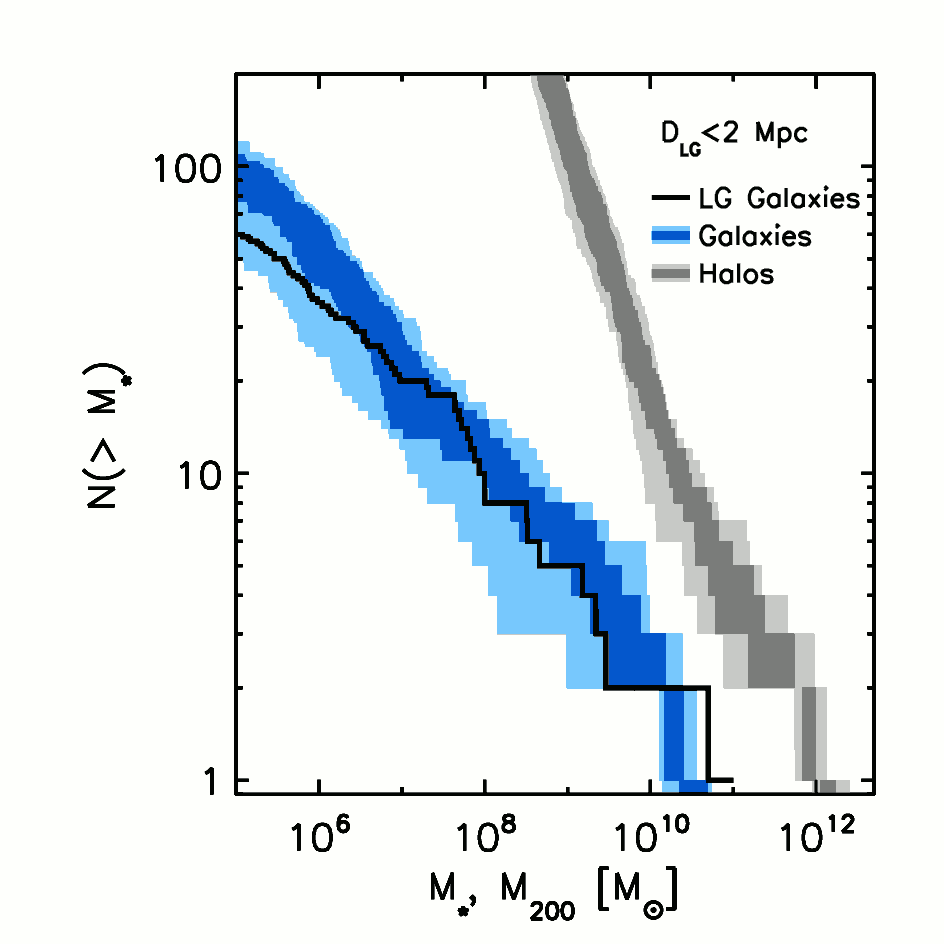}
\caption{Stellar mass functions from 12 {\sc Apostle} simulations at
  resolution L2 compared to observations. In the left and centre,
  shaded regions show the mass functions of satellites within 300 kpc
  of each of the primary (left) and secondary (centre) of the two main
  Local Group galaxies from each simulation volume, while lines show
  the observed stellar mass function within 300 kpc of M31 (left) and
  the MW (centre). In the right, the shaded region shows all galaxies
  within $2$~Mpc of the Local Group barycentre in the simulations,
  while the line is the stellar mass function of all known galaxies
  within the same region. On each panel, the dark colour-shaded areas
  bound the 16th and 84th percentiles; light shaded areas indicate the
  full range among our twelve Local Group realisations. For
  comparison, the grey area on each panel corresponds to the mass
  function of all dark matter halos. All observational data are taken
  from the latest compilation by \protect
  \cite{McConnachie-2012}. Note that while the M31 satellite count is
  likely to be complete to $10^5\Ms$, the count of satellites of the
  MW and the total count within $2$~Mpc should be considered as lower
  limits to the true numbers due to the limited sky coverage of local
  galaxy surveys and the low surface brightness of dwarf galaxies. See
  Fig.~\ref{fig:convergence} for numerical
  convergence.  \label{fig:smf}}
\end{center}
\end{figure*}

\section{Results}\label{sec:Results}

\subsection{Formation of LG galaxies}
While the Local Group observations are all made at $z=0$, and the
focus of our paper is on the relation between the observable stellar
component and the underlying dark matter model, our simulations allow
us to follow its evolution from before the formation of the first
stars to the present day. Fig.~\ref{fig:evolution} illustrates the
evolution of gas and stars in a comoving region of side length $200
\times 150$~h$^{-1}$ kpc from $z=12$ to $0$, centred on the particles
that become one of the central galaxies in one of the {\sc Apostle}
simulations at resolution level L1. At early times, the gas traces the
filamentary structure, and stars begin to form in the highest density
regions, often found in the nodes at the intersection of filaments. In
this simulation, the first stars begin to form at $z\sim17$, in the
progenitors of what will become the pair of main LG galaxies,
analogues to the Milky Way and M31. Immediately after the first stars
have formed, feedback associated with star formation begins to blow
out gas from the then very low mass dark matter and gas halos. At
$z=11.5$, reionization heats the intergalactic gas and rarefies gas
already collapsed in halos, quenching further gas cooling and
accretion into small halos. As a result, the formation of new galaxies
is disrupted, until sufficiently massive halos begin to form. Over
time, star formation begins anew in more and more halos, while
individual star-forming regions merge to assemble larger galaxies.

Shortly after $z\sim3$, the proto-galaxy undergoes a final major
merger, with minor mergers continuing up to $z=0$. The progenitor
continues to accrete new satellites that mostly lose their gas on
infall due to ram-pressure stripping. A stellar halo also builds up,
with shell-like and stream-like substructures originating from tidally
disrupted satellite galaxies. By $z=0$, a pair of large disk galaxies
have formed, both surrounded by shells and streams, along with many
dwarf galaxy satellites.

\subsection{Galaxies that only scratch the
  surface}\label{sec:results:introduction}
Fig.~\ref{FigLGImage} illustrates the dark matter and starlight in
another of our resimulations at redshift $z=0$. The top left panel
shows the dark matter distribution in a cube of side length 4 Mpc,
encompassing the spherical volume commonly considered as the Local
Group. It reveals a cosmic filament that envelopes the two principal
halos and most of the galaxies in the region. The bottom row zooms in
on a region of side length 2 Mpc around the simulated LG barycentre,
contrasting the distribution of dark matter (left panel) and star
light (right panel). While the simulations contains tens of thousands
of dark matter substructures, galaxies appear as highly biased tracers
of the dark matter, forming almost exclusively in the most massive
halos.

Also highlighted in the top left panel are the positions of the
satellite halos that host the eleven brightest satellites of one of
the central galaxies. The alignment of the satellites is indicative of
a thin plane seen in projection, that is also aligned with the
orientation of the filament.

The small insets in Fig.~\ref{FigLGImage} show the stellar structure
of some of the many galaxies formed in this simulation. The images use
multi-band colours rendered using a spectrophotometric model
\citep{Trayford-2015}. A variety of disk and spheroid morphologies,
luminosities, colours, and sizes are clearly visible, reminiscent of
the diversity of observed LG galaxies.

\subsection{No missing satellites}\label{sec:Results-SMF}
Fig.~\ref{fig:smf} shows the galaxy stellar mass functions in the
simulations, using data from all twelve of the {\sc Apostle} volumes
at resolution L2. Results are plotted both within $300$ kpc from each
of the two main galaxies per volume (labelled ``primary'' and
``secondary'' in order of halo mass), as well as within $2$ Mpc from
the LG barycentre, which includes both central and satellite galaxies.

The primary and secondary galaxies have $\tol{20}{10}{6}$ and
$\tol{18}{8}{5}$ satellites more massive than $M_*=10^5\Ms$ inside
$300$ kpc, respectively, where the errors indicate the scatter
equivalent to $1\sigma$ about the median values. This is in good
agreement with the observed number of MW and M31 satellites. Within
$2$ Mpc of the LG barycentre, there are $\sim60$ galaxies with
$M_*>10^5\Ms$ presently known; our simulations produce
$\tol{90}{20}{15}$. The modest number of luminous galaxies is in stark
contrast to the very large number of dark matter halos found within
the same volume, indicated by the grey shaded area in
Fig.~\ref{fig:smf}. While feedback from supernovae and stellar winds
regulates star formation in those halos where a dwarf galaxy has
formed, reionisation has left most of the low mass halos completely
dark. The observed stellar mass function of the LG and those of the MW
and M31 satellites are within the 1 $\sigma$ scatter of the average
stellar mass function in our resimulations over most of the stellar
mass range. The relative scatter is larger for the satellite galaxies,
reflecting the larger relative sampling error, and the fact that the
relative variation in single-halo mass among the different {\sc
  Apostle} volumes is larger than that of the total LG mass.

Excluding substructures, the stellar masses of the Milky Way and M31
analogues in our simulations lie in the range $1.5 - 5.5 \times
10^{10} \Ms$, on the low end compared to the observational estimates
for the Milky Way ($5 \times 10^{10} \Ms$ \citep{Flynn-2006,
  Bovy-2013}) but lower than those for M31 ($10^{11} \Ms$
\citep{Tamm-2012}). As noted by \cite{Schaye-2014}, the subgrid
physics used in the Reference model of the \eagle code, which we have
adopted in this work, generally results in slightly low stellar masses
in halos of around $10^{12}\Ms$ compared to abundance matching
expectations \citep[e.g.]{Guo-2010}, while the Milky Way and M31 both
appear to lie above the average stellar-to-halo mass relation. While
the predicted abundance of satellites and dwarf galaxies within the
Local Group depends on its total mass, as discussed in
Section~\ref{sec:Methods-ICs}, and in more detail by
\cite{Fattahi-2015}, we have selected our Local Group analogues based
on their dynamical properties in a pure dark matter simulation, and
independently of the stellar mass in the primaries, which may be
affected by the limitations of subgrid physics model.

That the simulations reproduce the stellar mass function of galaxies
and satellites in the LG over all resolved mass scales is remarkable,
given that these simulations use the very same \eagle model that
matches the shape and normalisation of the galaxy stellar mass
function in large cosmological volumes. Not only are our simulations
free of the ``missing satellites'' problem, but they indicate that the
observed stellar mass functions of the LG volume and of the MW and M31
satellites are entirely consistent with $\Lambda$CDM.

\begin{figure}
\begin{center}
 \resizebox{8.7cm}{!}{\includegraphics[trim={3.3cm 0 0 5.0cm},clip]{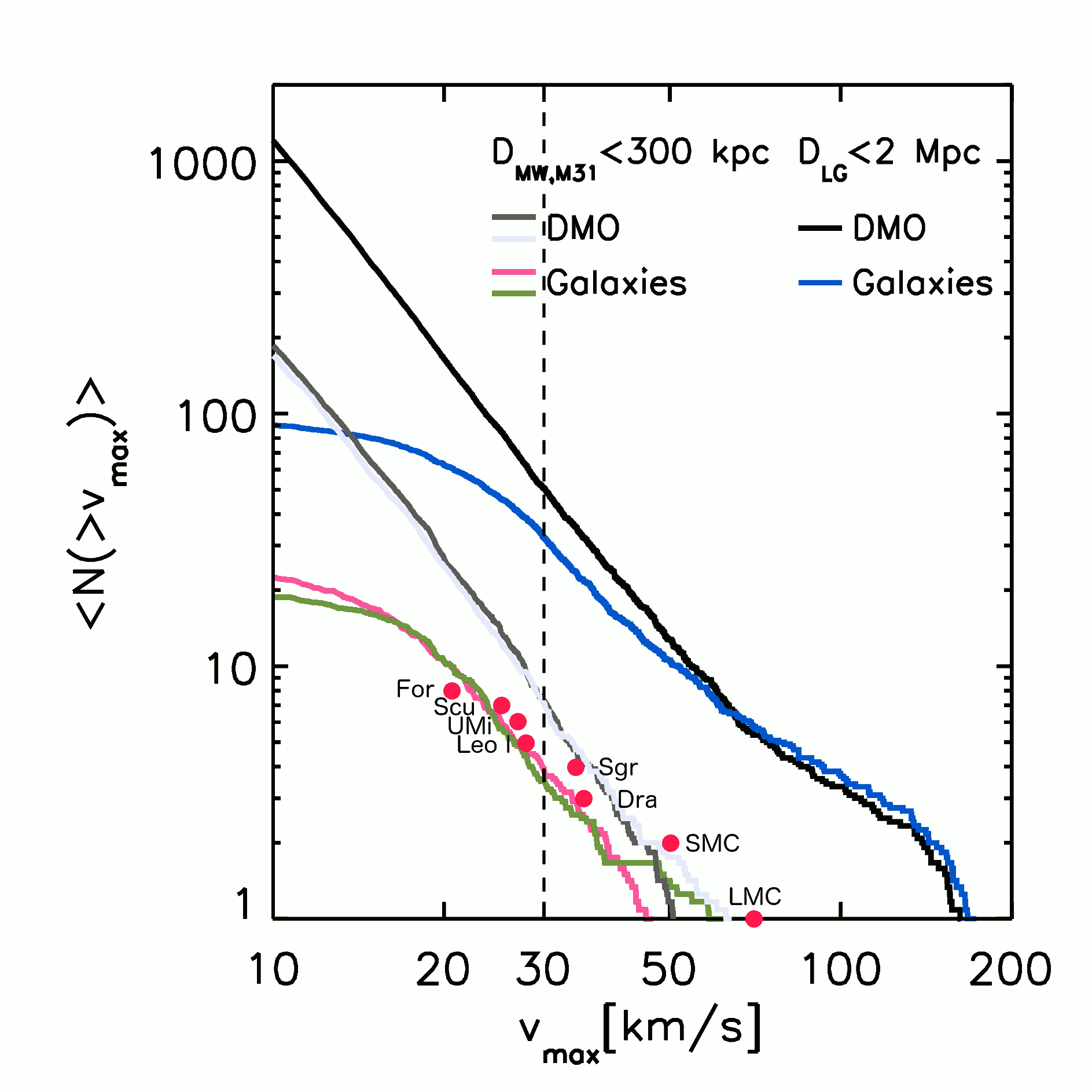}}%
 \caption{Cumulative number of halos as a function of maximum circular
   velocity, $\vm$, averaged over 12 {\sc Apostle} volumes at
   resolution level L2. The four bottom curves correspond to satellite
   halos within $300$ kpc of each of the two main galaxies; the top
   two curves to all systems within $2$~Mpc from the LG barycentre.
   Grey/black curves are from dark matter only (DMO) simulations.
   Coloured curves are for systems that contain luminous galaxies in
   the hydrodynamic runs. Red circles show measurements of the most
   massive MW satellites by \protect \cite{Penarrubia-2008}. For
   guidance, the dashed line denotes a $\vm$ value of
   $30$km/s. The abundance of satellites with $\vm >30$km/s is
   halved in the hydrodynamic simulations and matches the Milky Way
   observations. At lower values of $\vm$, the drop in the
   abundance relative to the DMO case increases as fewer and fewer
   subhalos host an observable galaxy. See Fig.~\ref{fig:convergence}
   for numerical convergence.}\label{FigVmaxFunc}
\end{center}
\end{figure}

\subsection{The Baryon Bailout}\label{sec:Results-TBTF}
We next consider the ``too-big-to-fail'' problem
\citep{Boylan-Kolchin-2011, Parry-2012}. As demonstrated by
\cite{Strigari-2010} using the {\sc Aquarius} dark matter only (DMO)
simulations \citep{Springel-2008}, a Milky Way mass halo in
$\Lambda$CDM typically contains at least one satellite substructure
that matches the velocity dispersion profiles measured for each of the
five Milky Way dwarf spheroidal satellites for which high-quality
kinematic data are available. However, that work addressed neither the
question of whether those halos which match the kinematics of a
particular satellite would actually host a comparable galaxy, nor
whether an observed satellite galaxy can be found to match each of the
many predicted satellite halos. Indeed, the identification in the same
simulations of an excess of massive substructures with no observable
counterparts, and the implication that the brightest satellites of the
Milky Way appear to shun the most massive CDM substructures,
constitutes the ``too big-to-fail'' problem
\citep{Boylan-Kolchin-2011, Parry-2012}.  A simple characterisation of the
too-big-to-fail problem is given by the number of satellite halos with
maximum circular velocities, $\vm = \max \left(\sqrt{G M(<r) / r}
\right)$, above $\sim 30$ km/s, where all satellite halos are expected
to be luminous \citep{Okamoto-2008, Sawala-2014b}. Only three MW
satellites are consistent with halos more massive than this limit (the
two Magellanic Clouds and the Sagittarius dwarf), whereas {\it dark
  matter only} (DMO) $\Lambda$CDM simulations of MW-sized halos
produce two to three times this number.

\begin{figure}
\begin{center}
\includegraphics[width=8.6cm, trim={3.cm 0 2.cm 1.5cm},clip]{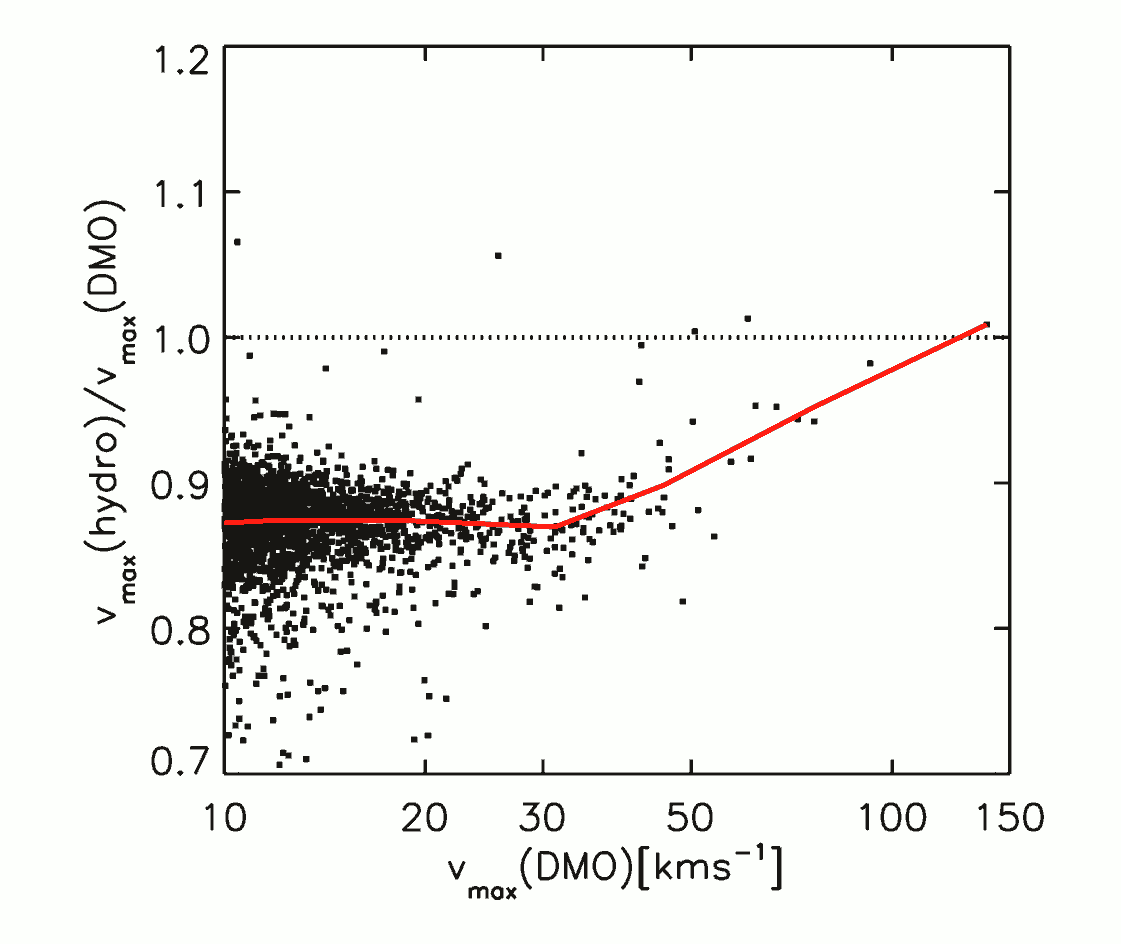}
\caption{Ratio between the maximum circular velocity, $\vm$, of
  individual, isolated halos in the hydrodynamic simulation and the
  DMO simulation of the same volume, as a function of $\vm$, at
  resolution L1. The red line shows the binned median ratio. The loss
  of baryons and the truncated growth leads to a reduction in $\vm$
  for halos below $\vm \sim 100$km/s. \label{fig:reduction} }
\end{center}
\end{figure}

Indeed, as shown in Fig.~\ref{FigVmaxFunc}, when we consider the DMO
counterparts of our LG simulations, the MW and M31 halos each contain
an average of $7-8$ satellites with $\vm >30$ km/s inside 300 kpc,
more than twice the observed number of luminous satellites. This is
despite the fact that, in order to match the most recent dynamical
constraints \citep{Gonzalez-2013, Penarrubia-2014}, the average halo
masses of M31 and the MW in the {\sc Apostle} simulations are lower
than those in which the problem was first identified
\citep{Boylan-Kolchin-2011}.

\begin{figure}
\begin{center}
\includegraphics[trim=0cm 0.7cm 0 0,clip=true, height=3.8cm]{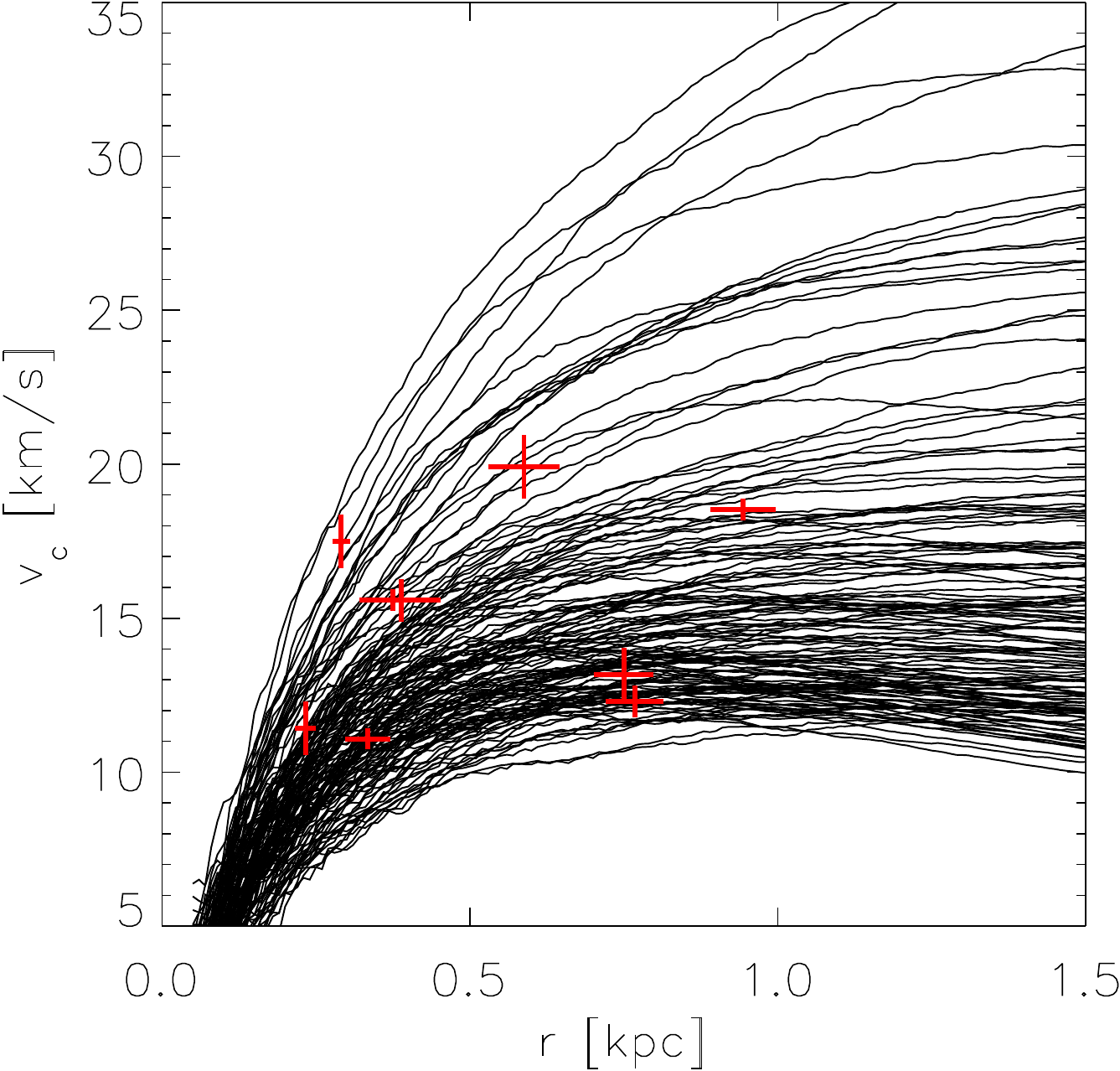}
\includegraphics[trim=1cm 0.7cm 0 0,clip=true, height=3.8cm]{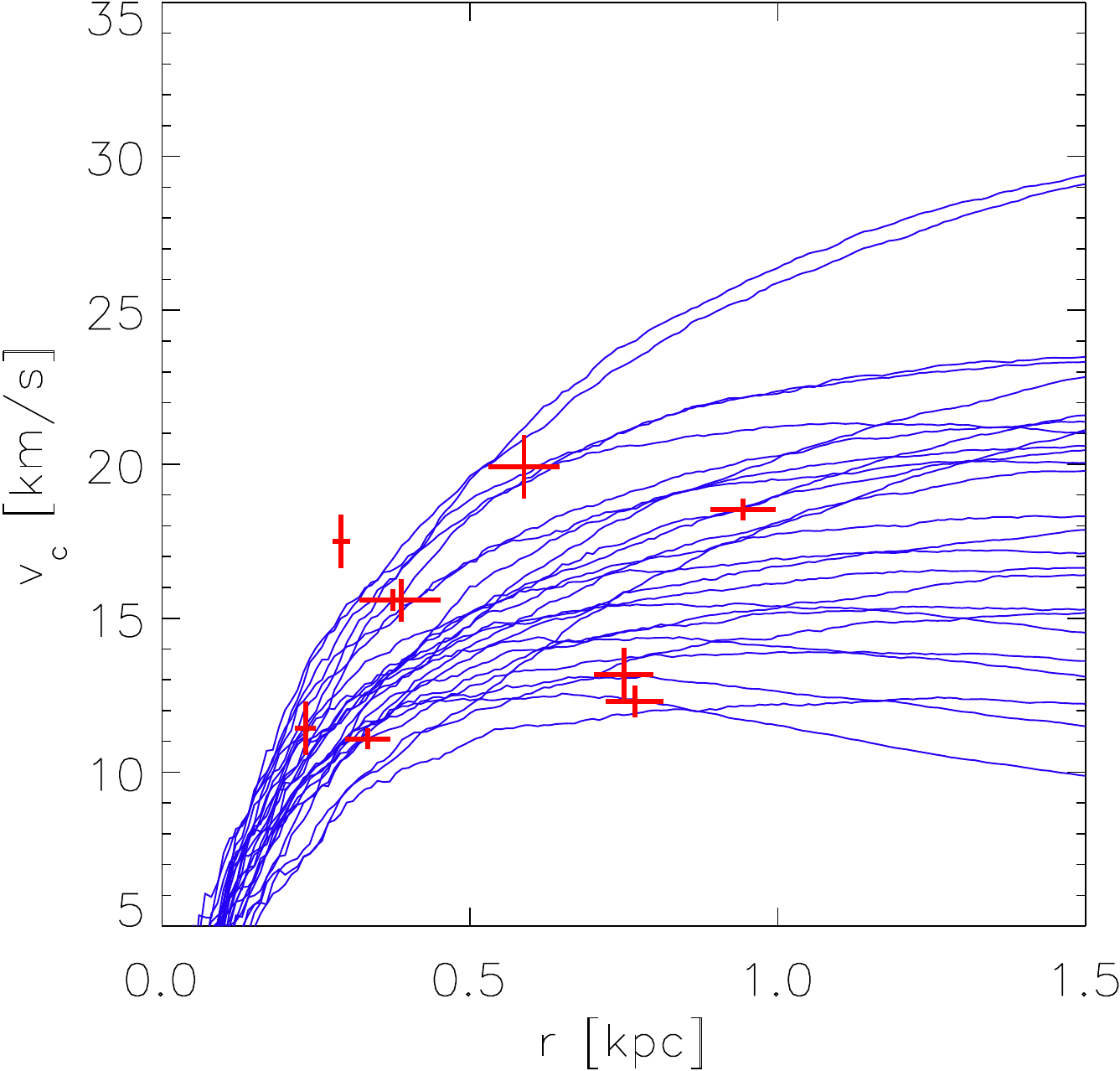} \\
\includegraphics[trim=0cm 0.7cm 0 0,clip=true, height=3.8cm]{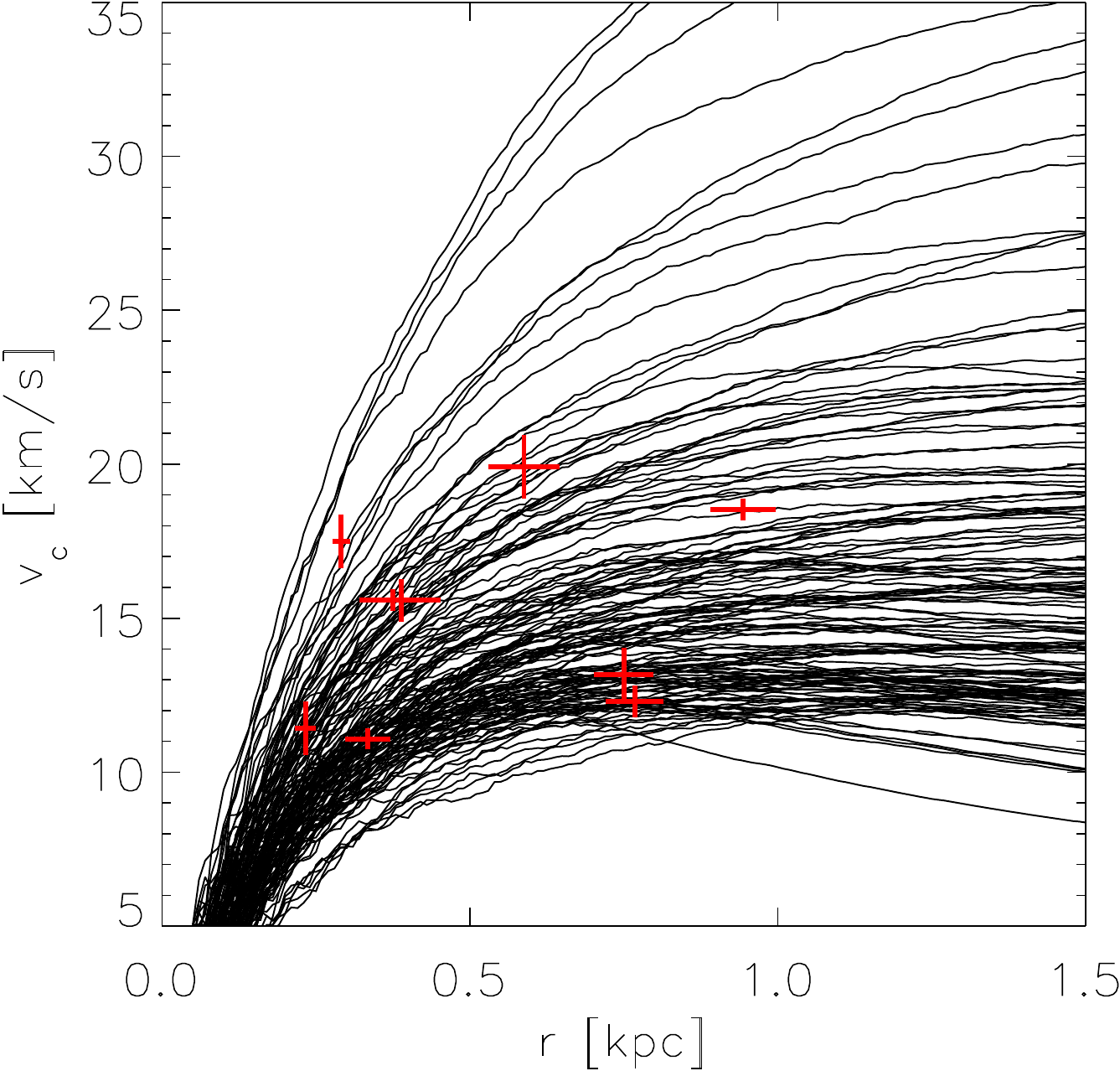}
\includegraphics[trim=1cm 0.7cm 0 0,clip=true, height=3.8cm]{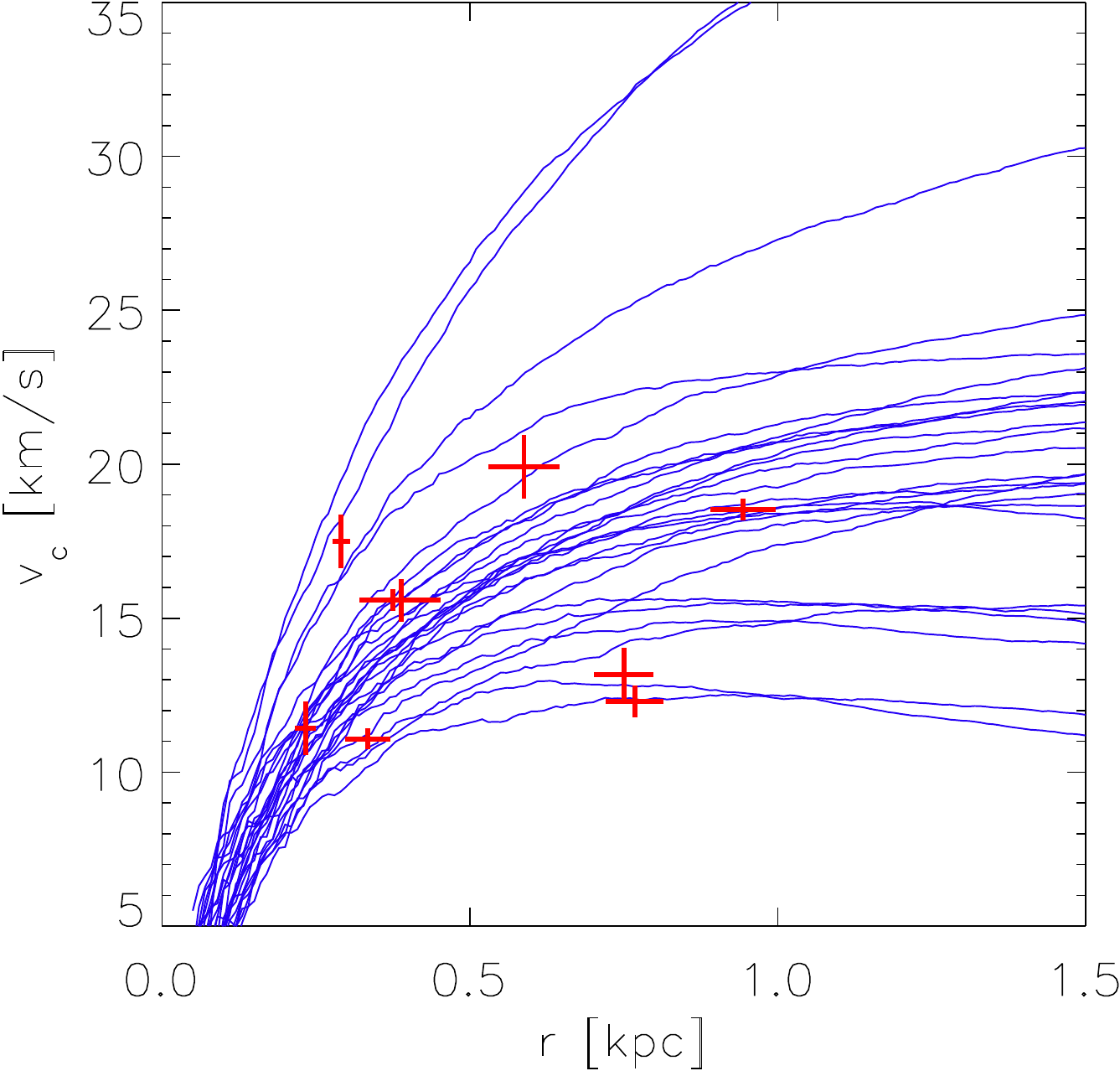} \\
\includegraphics[trim=0cm 0.7cm 0 0,clip=true, height=3.8cm]{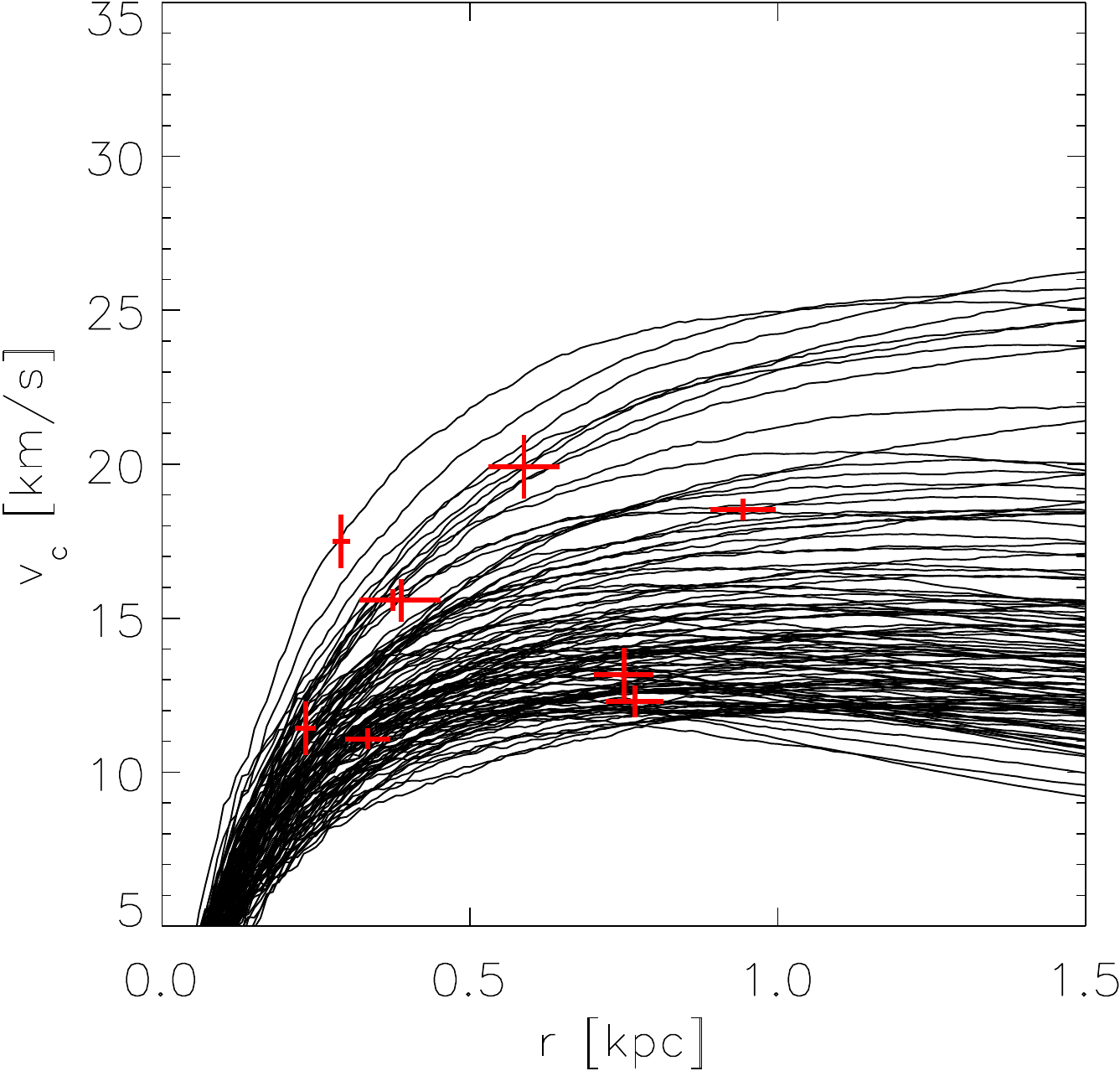}
\includegraphics[trim=1cm 0.7cm 0 0,clip=true, height=3.8cm]{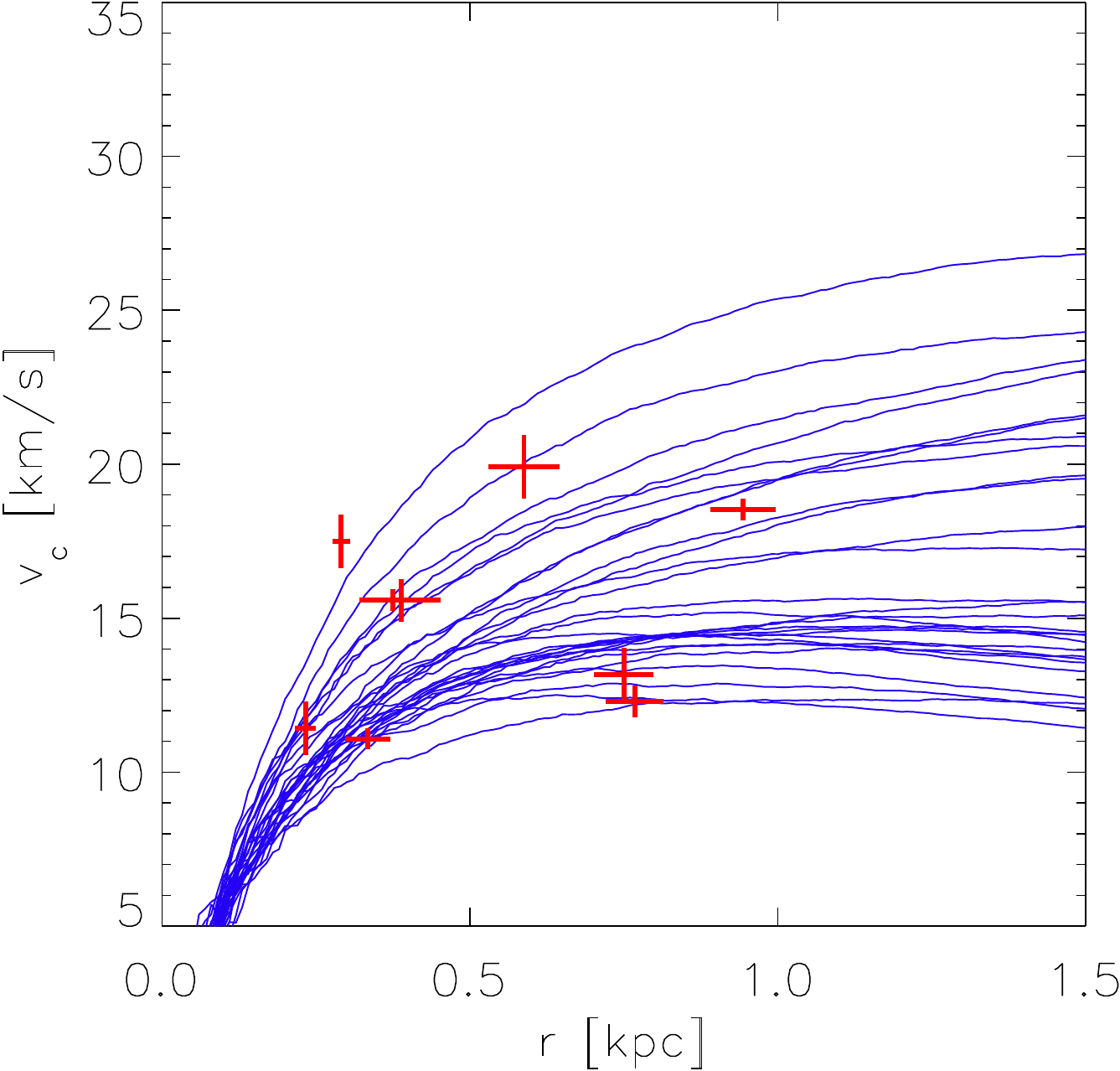} \\
\includegraphics[trim=0cm 0.7cm 0 0,clip=true, height=3.8cm]{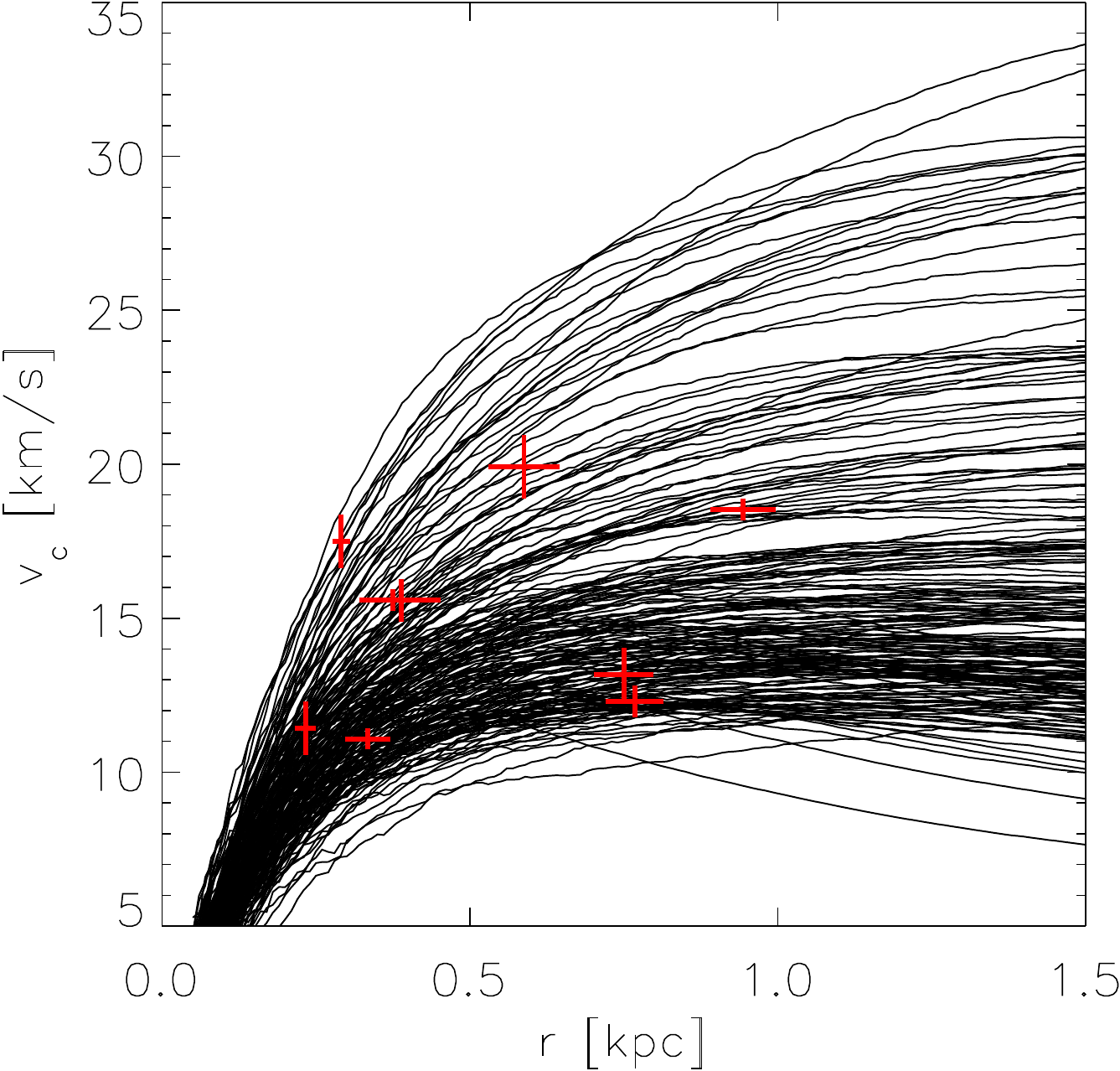}
\includegraphics[trim=1cm 0.7cm 0 0,clip=true, height=3.8cm]{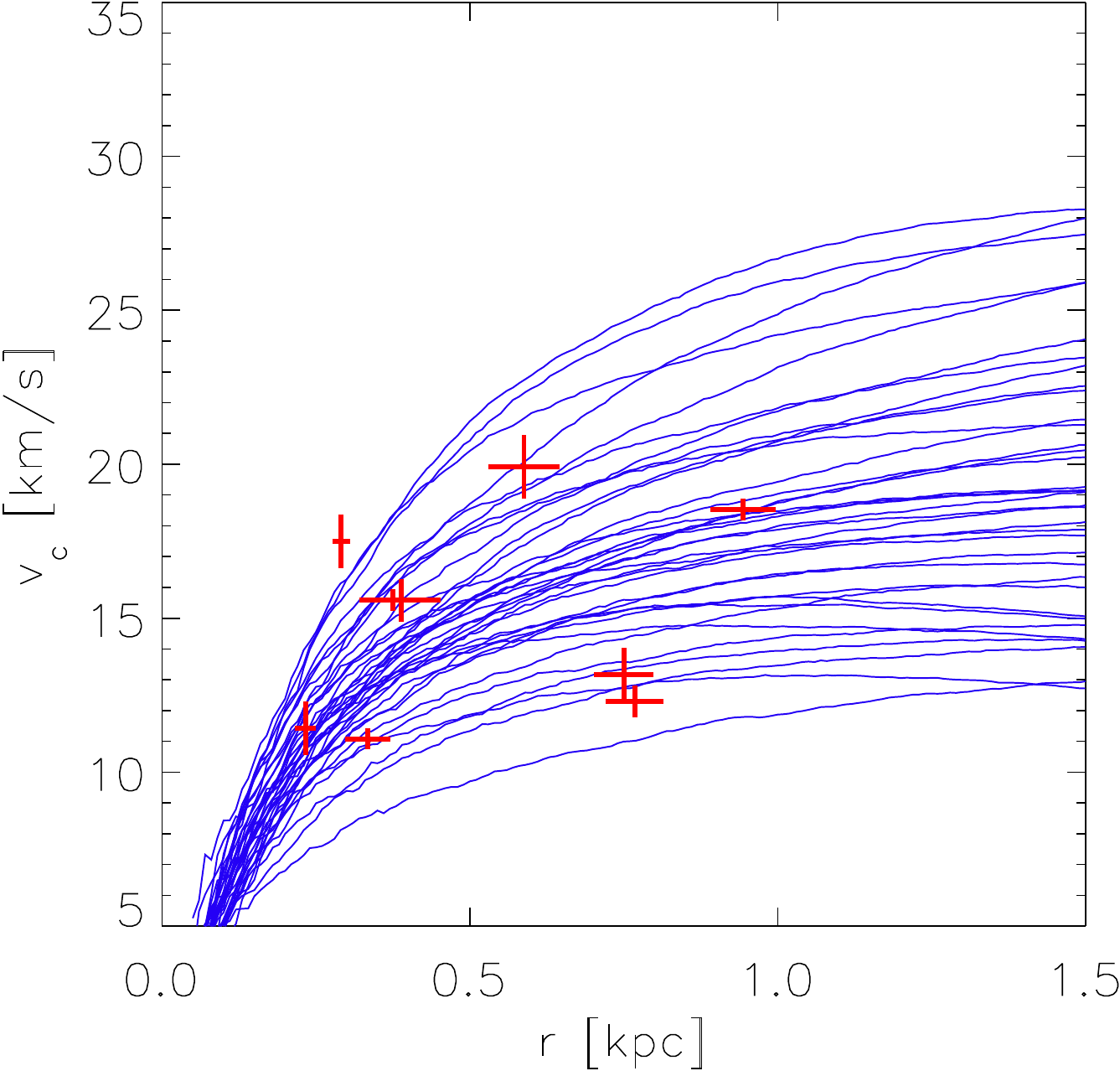}

\caption{Left: circular velocity profiles of all satellites with $\vm
  > 12$~km/s and within 300 kpc from the halo centre, in four
  individual LG halos from two {\sc Apostle} simulations at resolution
  L1. Right: as above, for the satellite galaxies in the corresponding
  hydrodynamic simulations. Overplotted in red on each panel are the
  half-mass radii and corresponding circular velocities of nine
  observed MW dwarf spheroidal satellites adopted from \protect
  \cite{Wolf-2010}. While three of the four halos in the DMO
  simulations have multiple massive subhalos without observable
  counterparts, the discrepancy is resolved in the hydrodynamic
  simulations. As the LMC, SMC and the Sagittarius dSph, all
  consistent with $\vm > 30$ km/s, are excluded from this sample, we
  have also removed the three satellites with the highest $\vm$ values
  from each panel.} \label{fig:rotation-curves-hydro-dmo}
\end{center}
\end{figure}

The situation changes, however, when we consider the {\it
  hydrodynamic} simulations: each main galaxy in our hydrodynamic
simulation has on average only $3-4$ luminous satellites with $\vm
>30$~km/s. Furthermore, the average velocity function of the most
massive substructures across the {\sc Apostle} simulations appears to
be in excellent agreement with the MW estimates quoted by
\cite{Penarrubia-2008} and overplotted as red circles in
Fig.~\ref{FigVmaxFunc}. It should be noted that the true $\vm$ values
of dwarf spheroidal galaxies cannot easily be measured, and their
estimates rely on additional assumptions. For Fig.~\ref{FigVmaxFunc},
we have chosen to use measurements obtained independently from our own
simulations, but we revisit this topic in Section~\ref{sec:vmax} where
we provide ranges of the likely $\vm$ values of nine MW dwarf
spheroidals with measured stellar masses and velocity dispersions.

\begin{table*}

\begin{center}
  \caption{Structural parameters and $\vm$ estimates for Milky-Way
    satellite galaxies with M$_* > 10^5 \Ms$ \label{table:vmax}} 
\centering 
\begin{tabular}{lcccccccc}
\hline
\hline  
& $M_*$ & $\rh$ & $\sqrt{\left<\sigma^2_{los}(\rh)
  \right>} $&$\vm$& $\vm$& $\vm$& $\vm$ (DMO) & $\vm$ (hydro)  \\ 
  &  $ [\Ms]$ & [pc] & [km/s] & [km/s] &  [km/s] & [km/s]&  [km/s]
  & [km/s]\\
\hline 
Notes and References & (1) & (2) & (2) & (3) & (4) & (5) & this work
(6) & this work (7) \\
\hline 

Carina & $4.3^{+1.1}_{-0.9}\times10^5$ & $334 \pm 37$ &  $6.4 \pm0.2$&  16 & 17 & $11.4 ^{+1.1}_{-1.0}$  &  $13.7^{+4.8}_{-2.2} $ & $14.6 ^{+6.4}_{-2.9}$ \\
Draco  & $2.2^{+0.7}_{-0.6} \times 10^5$ & 291 $\pm$ 14  & 10.1 $\pm$0.5 & -- & 35&$20.5 ^{+4.8}_{-3.9}$  &  $27.6^{+12}_{-4.1}$ & $23.7^*$ \\
Fornax & $1.7^{+0.5}_{-0.4} \times 10^7$ & 944 $\pm$ 53  & 10.7$\pm$0.2 & 21 & 21& $17.8 ^{+0.7}_{-0.7}$ &  $19.6^{+1.4}_{-1.1}$ &$20.4^*$ \\
Leo I   & $5.0^{+1.8}_{-1.3} \times 10^6$ & 388 $\pm$ 64  & 9.0 $\pm$0.4 &  22 & 30& $16.4 ^{+2.3}_{-2.0}$ & $ 18.5^{+9.0}_{-3.4}$ & $18.8^{+12.3}_{-4.0}$ \\
Leo II   & $7.8^{+2.5}_{-1.9} \times 10^5$ & 233 $\pm$ 17  &  6.6 $\pm$ 0.5 &  -- & 18&$12.8 ^{+2.2}_{-1.9}$ &  $14.2^{+6.4}_{-2.6}$ & $17.7^{+6.0}_{-3.2} $\\
Sculptor   &  $2.5^{+0.9}_{-0.7} \times 10^6$ & 375 $\pm$ 54  & 9.0 $\pm$ 0.2 & 26  & 27&$17.3 ^{+2.2}_{-2.0}$ & $20.9^{+9.6}_{-4.0}$ & $20.1^{+7.2}_{-1.6} $\\
Sextans  & $5.9^{+2.0}_{-1.4} \times 10^5$ & 768 $\pm$ 47 & 7.1 $\pm$ 0.3 &  12 & 11&$11.8 ^{+1.0}_{-0.9}$ &$12.4^{+1.6}_{-1.0}$ & $13.2^{+2.5}_{-1.5}$ \\
Ursa Minor  &$3.9^{+1.7}_{-1.3} \times 10^5$ & 588 $\pm$ 58  & 11.5 $\pm$ 0.6 &  -- & 29&$20.0 ^{+2.4}_{-2.2}$ & $22.6^{+6.6}_{-4.0}$ & $21.5^{+2.8}_{-0.4} $\\
CanVen I  & $2.3^{+0.4}_{-0.3} \times 10^5$ & 750 $\pm$ 48& 7.6 $\pm$ 0.5 &  -- & --&$11.8^{1.3}_{-1.2}$ & $12.8^{+2.8}_{-1.6}$ & $13.1^{+3.4}_{-1.4}$ \\
\hline  \\
\end{tabular}
\end{center}

(1): \protect \cite{McConnachie-2012}, (2): \protect \cite{Wolf-2010},
(3): \protect \cite{Strigari-2010}, (4) \protect \cite{Penarrubia-2008}, (5) \protect \cite{Boylan-Kolchin-2012}, (6) satellite halos from DMO APOSTLE
simulations that match the observed $\vh$ at the observed
$\rh$, (7) satellite halos from hydrodynamic {\sc Apostle}
simulations that match the observed $\vh$ at the observed
$\rh$, and that host galaxies that match the observed stellar
mass. $^*$For Fornax and Draco, there are too few simulated counterparts to
estimate the range reliably. 
\end{table*}

\begin{figure*}
\begin{center}
\includegraphics[trim=2cm 18cm 8cm 0.5cm,clip=true,width=8.5cm]{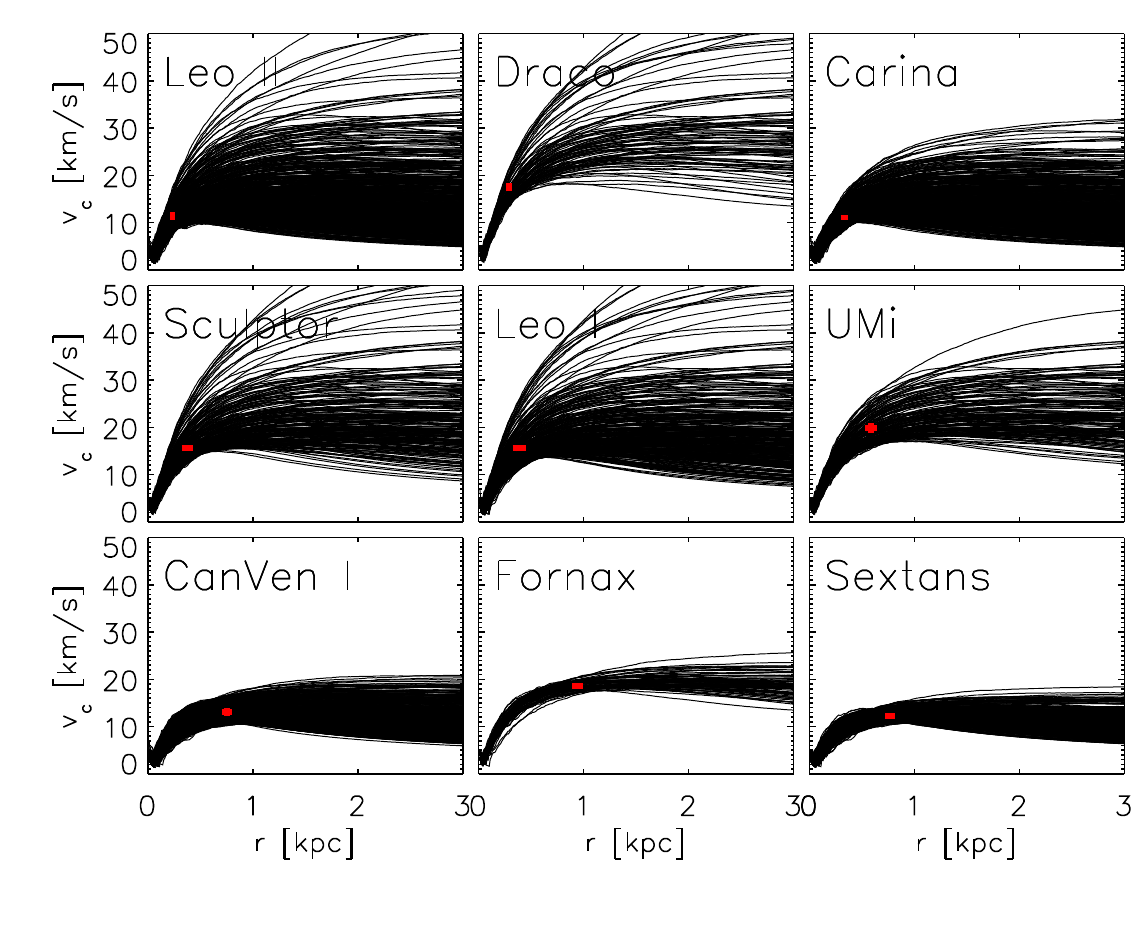}
\includegraphics[trim=2cm 18cm 8cm 0.5cm,clip=true,width=8.5cm]{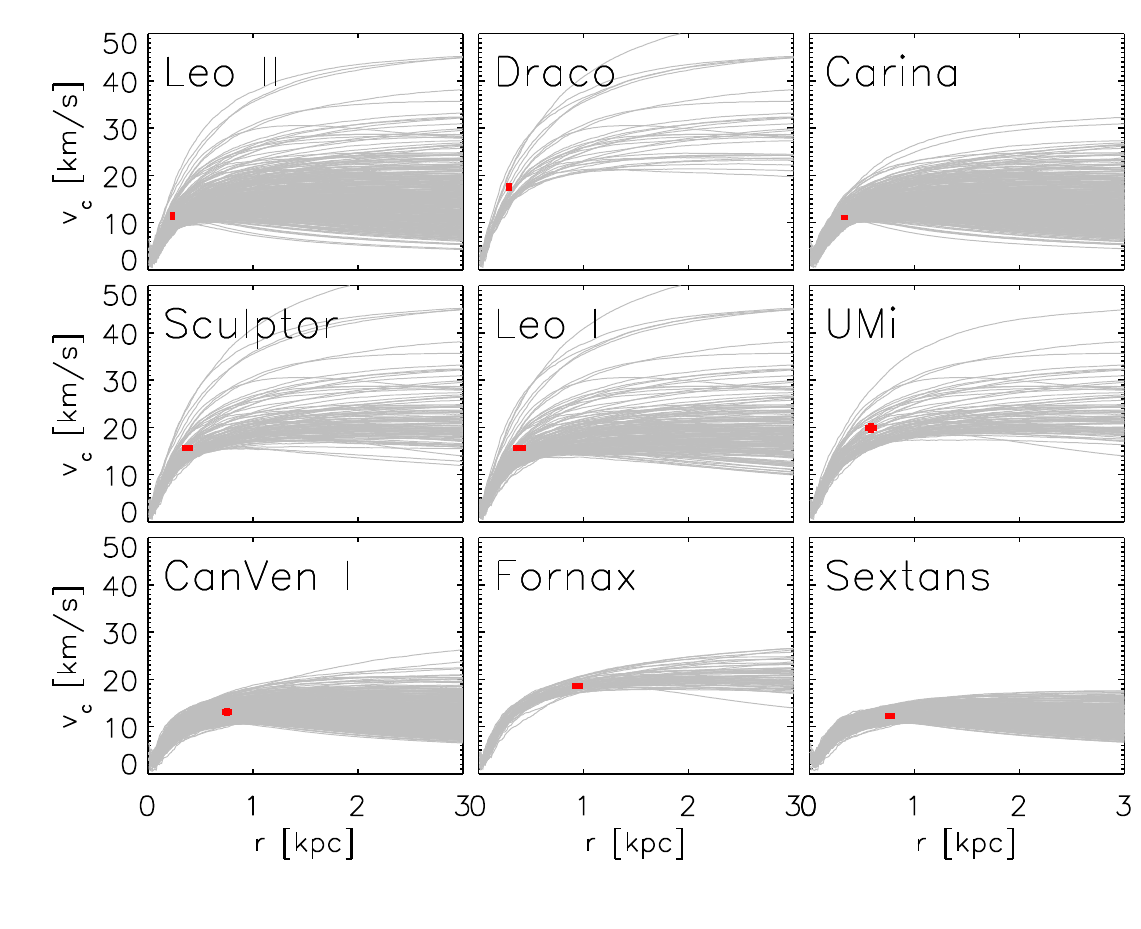}
\\
\vspace{-0.5cm}
\includegraphics[trim=2cm 18cm 8cm 0.5cm,clip=true,width=8.5cm]{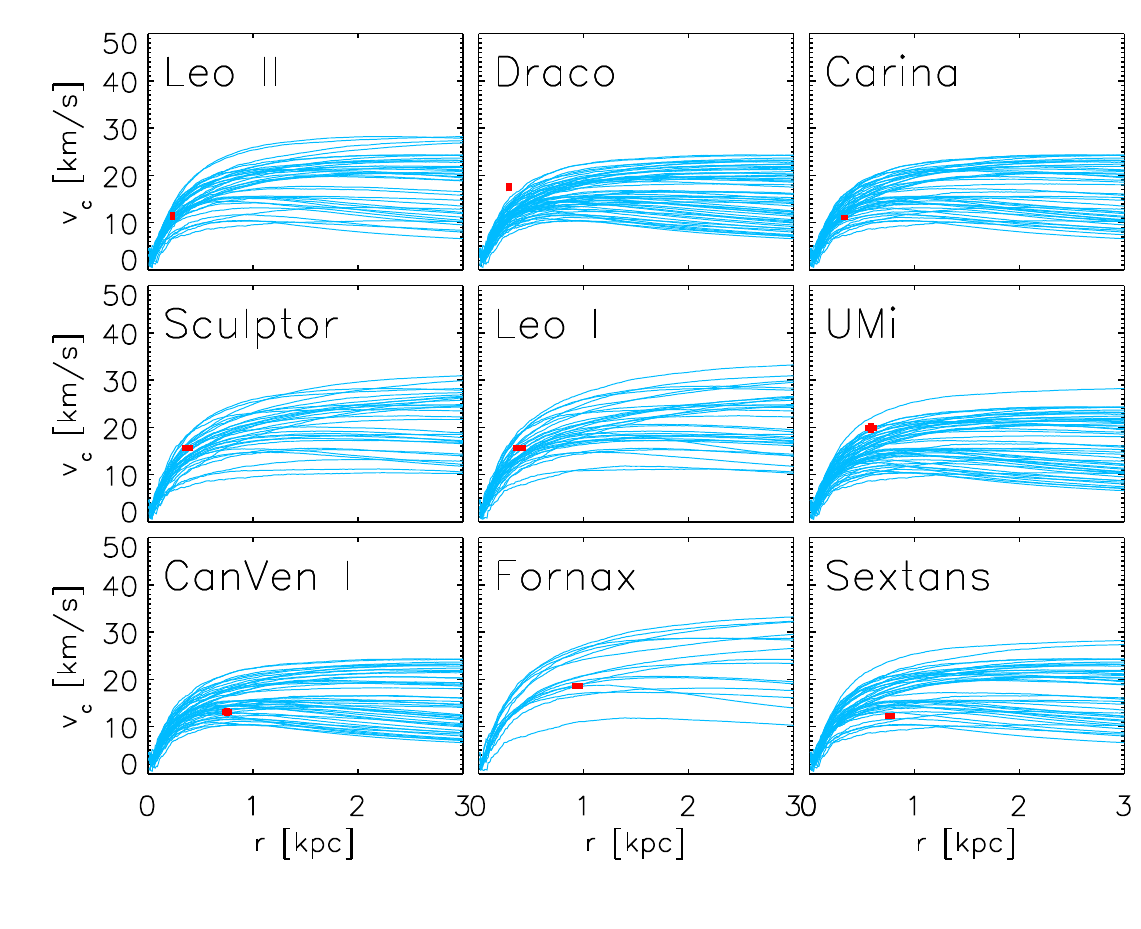}
\includegraphics[trim=2cm 18cm 8cm 0.5cm,clip=true,width=8.5cm]{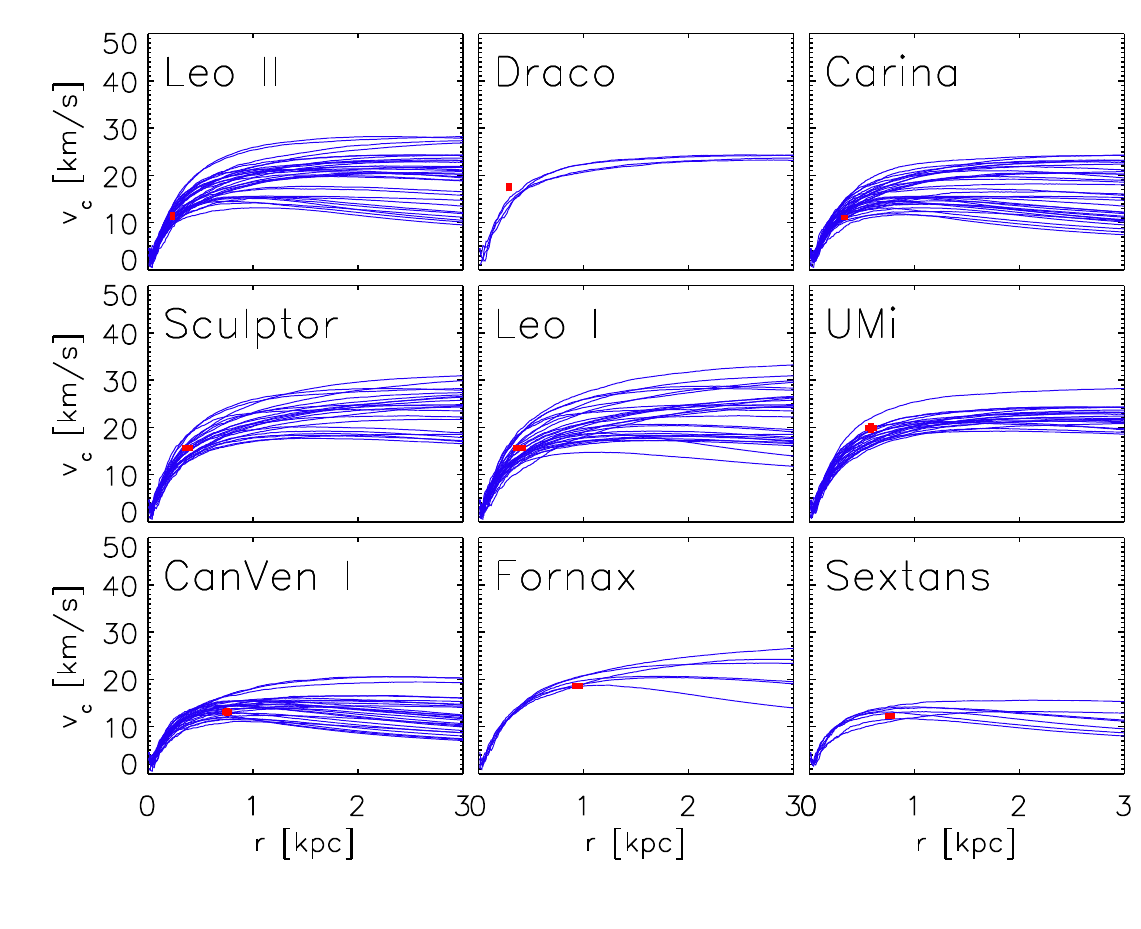}

\caption{Circular velocity profiles for satellites from a total of
  four halos from two of our LG simulations at resolution L1, which
  have been matched to 9 observed dSph galaxies. On the top left,
  black lines show satellites from the DMO simulations whose
  circular velocities are within 3$\sigma$ of the value corresponding
  to the observed velocity dispersion at the observed half-light
  radius. On the top right, grey lines show satellites from the
  corresponding hydrodynamic simulations selected with the same
  criteria. On the bottom left, cyan lines are for satellites from
  the hydrodynamic simulations, selected by stellar mass to be within
  a factor of 2 of the observed value, and on the bottom right, dark blue
  lines are for satellites from the hydrodynamic simulation that
  satisfy both the stellar mass and circular velocity criteria. On all panels, red symbols show measurements
  for observed dwarf spheroidals at the projected 3D half-light
  radius, adopted from \protect \cite{Wolf-2010}. 
  \label{fig:rotation-curves} }
\end{center}
\end{figure*}

Several factors contribute to the reduction in the measured satellite
$\vm$ function in our hydrodynamic simulations compared to DMO
simulations: (i) a reduction in the mass of each subhalo due to
baryonic effects as discussed below, (ii) the failure of a fraction of
subhalos of $\vm < 30$~km/s to form any stars, and (iii) those halos
of $\vm < 30$~km/s today that actually contain observable dwarf
galaxies having been affected by tidal stripping, even more strongly
than typical satellite halos of the same mass today.

In Fig.~\ref{fig:reduction}, we compare the maximum circular velocity
of individual isolated halos matched in our hydrodynamic and DMO
simulations. In agreement with \cite{Sawala-2013} and
\cite{Schaller-2015}, we find that while the more massive halos of
$\vm > 100$~km/s such as those that host the MW and M31 are not
significantly affected by baryonic effects, the halos of dwarf
galaxies end up being less massive than their DMO counterparts,
because the loss of baryons due to reionization and supernova
feedback, results in a reduced halo growth rate and leads to a $\sim
15\%$ reduction in $\vm$. The average reduction in mass is similar for
the halos of satellite and isolated galaxies prior to infall, but the
more massive satellites in the hydrodynamic simulations experience a
further mass loss relative to their DMO counterparts due to the
ram-pressure stripping of the remaining gas.

For halos below 30~km/s, the intrinsic reduction in $\vm$ due
to baryonic effects is compounded by the fact that not all low-mass
halos host galaxies: at $10$ km/s the fraction of luminous systems is
well below $10\%$ and decreases even further towards lower masses.

Fig. ~\ref{fig:rotation-curves-hydro-dmo} shows a comparison of
individual circular velocity curves of the most massive satellites
within 300 kpc of four of the main LG galaxies, in the DMO simulations
(black, left column) and in the corresponding hydrodynamic (blue,
right column) simulations. While there is considerable scatter due to
the fact that individual satellites can evolve differently in the two
simulations, particularly after infall, it can be seen that the
satellites in the hydrodynamic simulations have systematically lower
circular velocity curves compared to the DMO counterparts. While three
of the four halos in the DMO simulations contain a number of
satellites whose circular velocity curves cannot be matched by any of
the observed satellite galaxies shown, the velocity curves of
satellites in the hydrodynamic simulations are consistent with the
observed stellar kinematics.

\subsection{$\mathrm{V}_\mathrm{max}$ estimates for dwarf spheroidal
  galaxies}
\label{sec:vmax} Because the visible stellar components of dwarf
spheroidal galaxies probe only the innermost part of their dark matter
halos, connecting the measured line-of-sight velocity dispersion,
$\sigma_{los}$, and the size of a satellite to the maximum circular
velocity, $\vm$, of its halo is not straightforward, and relies on
several assumptions.

\cite{Penarrubia-2008} estimated $\vm$ values of individual Milky Way
satellite galaxies, assuming that their stellar and dark matter
components follow King profiles and NFW profiles,
respectively. \cite{Boylan-Kolchin-2012} also gave values for $\vm$
for nine Milky-Way satellites with stellar masses above
$10^5\Ms$. Using the result of \cite{Wolf-2010} that the uncertainty
on the enclosed mass for an observed line-of-sight velocity dispersion
is minimal at the stellar half-light radius, $\rh$, they used the
circular velocity profiles of satellite halos from the {\sc Aquarius}
DMO simulations to determined the most likely $\vm$ value of satellite
halos in CDM that match the measured values of $\rh$ and
$\sigma_{los}(\rh)$. Independently, using the same simulations,
\cite{Strigari-2010} also determined the $\vm$ values of five dwarf
spheroidals with resolved kinematics. Instead of relying on the
velocity dispersion at the half-light radius, they determine the most
likely value of $\vm$ for a given observed satellite from the
best-fitting velocity dispersion profile in the simulated halos. The
observed structural parameters of nine individual Milky Way satellite
galaxies compiled by \cite{Wolf-2010}, and the $\vm$ estimates of
\cite{Strigari-2010}, \cite{Penarrubia-2008} and
\cite{Boylan-Kolchin-2012} are reproduced in Table~\ref{table:vmax}.

\begin{figure*}
\begin{center}

\begin{overpic}[width=8.4cm]{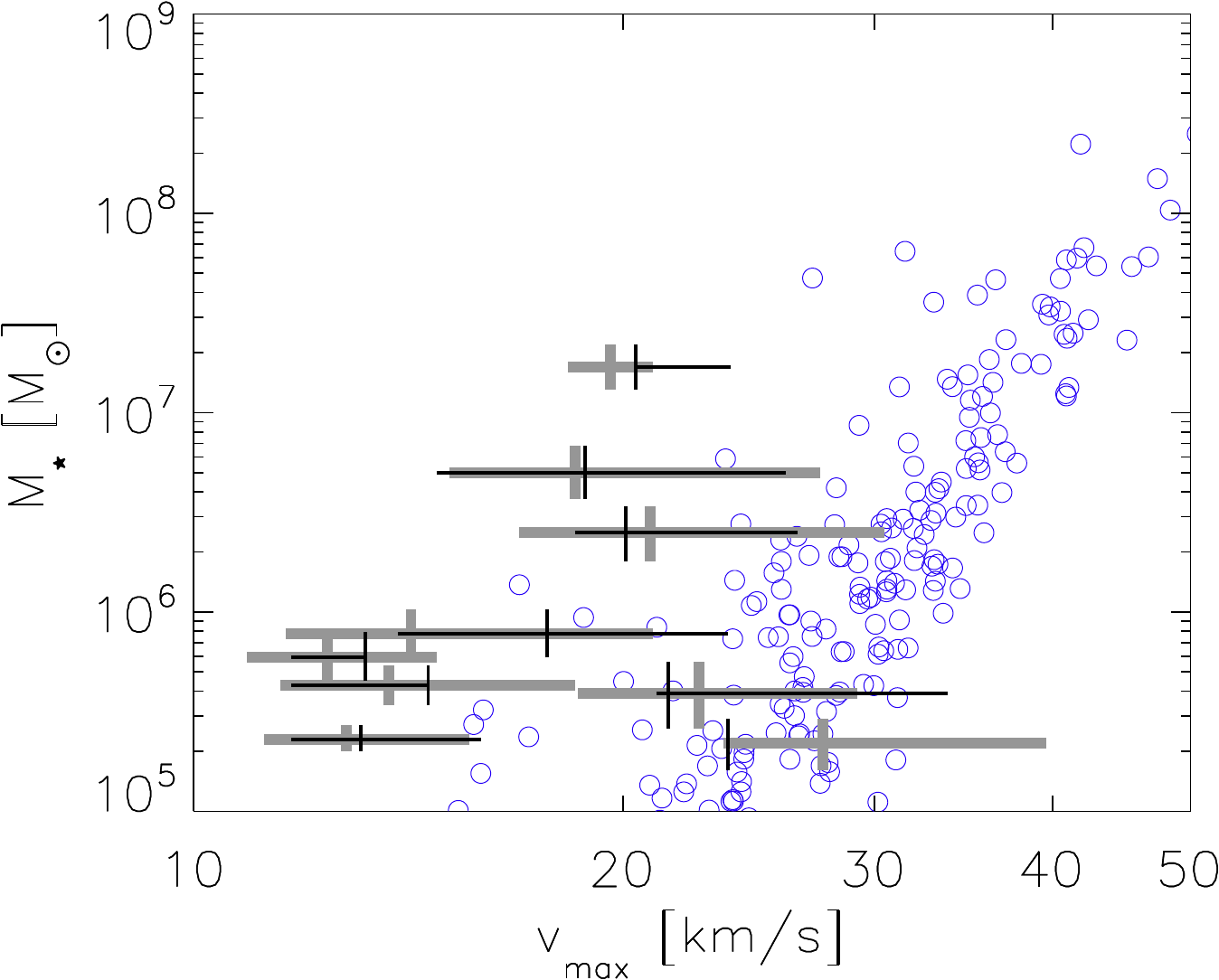}
\small
  \put(31,20){\color{red}Car}
  \put(61,16){\color{red}Dra}
  \put(49,53){\color{red}For}
  \put(43,45){\color{red}Leo I}
  \put(30,31){\color{red}Leo II}
  \put(50,32){\color{red}Scl}
  \put(18,25.5){\color{red}Sex}
  \put(52,25){\color{red}U Mi}
  \put(24,16){\color{red}CV I}
\end{overpic}
\begin{overpic}[width=8.4cm]{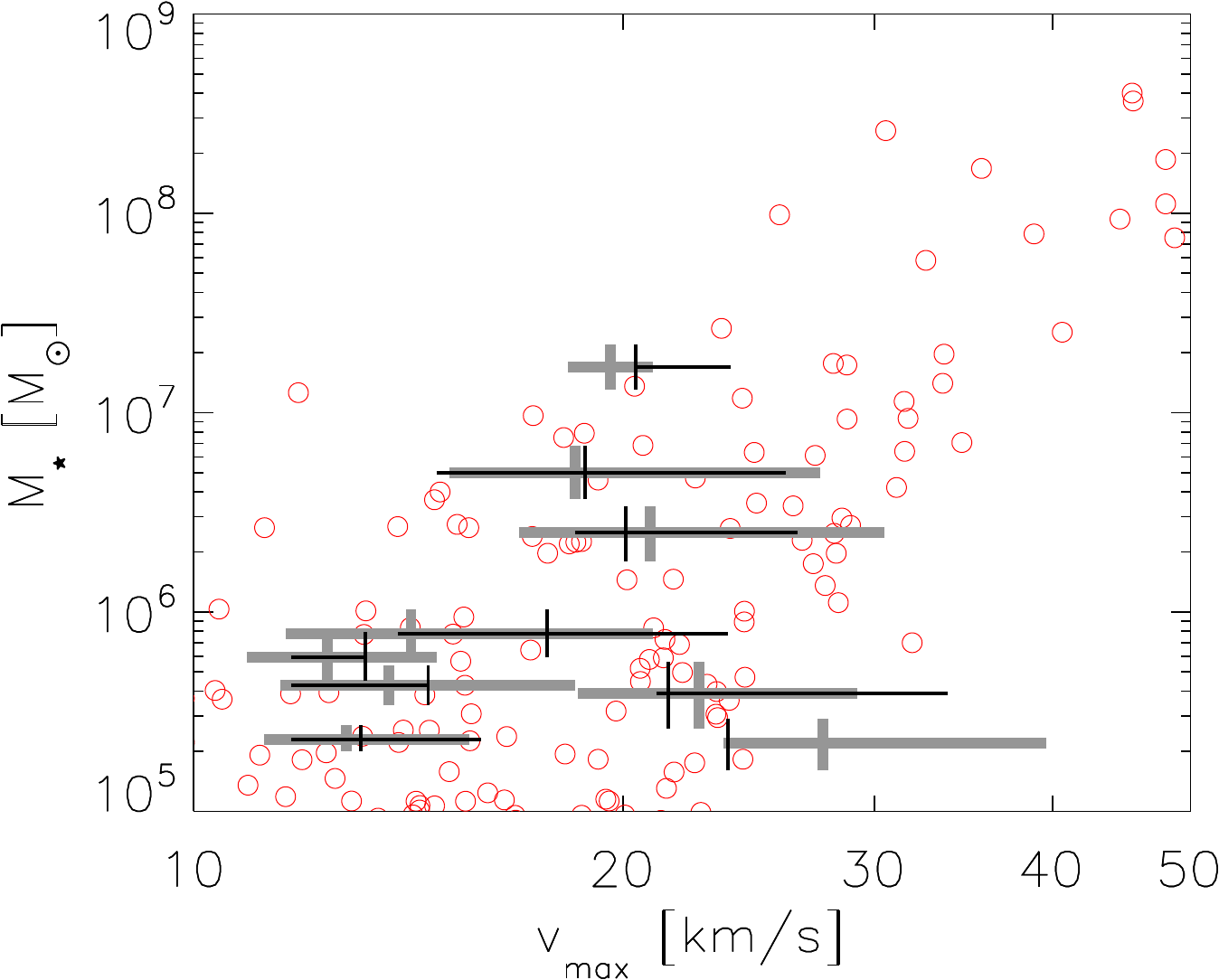}
\small
  \put(31,20){\color{blue}Car}
  \put(61,16){\color{blue}Dra}
  \put(49,53){\color{blue}For}
  \put(43,45){\color{blue}Leo I}
  \put(30,31){\color{blue}Leo II}
  \put(50,32){\color{blue}Scl}
  \put(18,25.5){\color{blue}Sex}
  \put(52,25){\color{blue}U Mi}
  \put(24,16){\color{blue}CV I}
\end{overpic}

\caption{Stellar mass as a function of maximum circular velocity,
  $\vm$, for isolated galaxies (blue circles, left) and satellite
  galaxies (red circles, right) from two LG simulations at resolution
  L1. Also shown are measurements for the nine MW satellites listed in
  Table~1, with stellar masses adopted from \protect \cite{Wolf-2010}
  and the range of possible $\vm$ values computed from the observed
  $\rh$ and $\sigma_{los}(\rh)$ as discussed in
  Section~\ref{sec:vmax}, based on halos from the DMO simulations
  (grey) and based on satellite galaxies from the hydrodynamic
  simulations (black). While isolated galaxies in the simulation fall
  below the MW satellites, the reduction in $\vm$ particularly of
  low-mass satellites by tidal stripping, brings the simulated
  satellite galaxies into good agreement with observations.
  \label{fig:mstar-mhalo} }
\end{center}
\end{figure*}

For those galaxies where multiple estimates are available, the $\vm$
values of \cite{Boylan-Kolchin-2012} tend to be lower than those
obtained by the two other authors. Indeed, as pointed out by
\cite{Pawlowski-2015}, these low values would still lead to an
overprediction of the MW satellite $\vm$ function in our simulations,
even though the extrapolation down to 10 km/s is somewhat
misleading, because it only includes $\vm$ values for a small fraction
of the known MW satellites.

However, the most likely subhalo in a DMO simulation like {\sc
  Aquarius} whose enclosed mass inside $\rh$ corresponds to the
observed $\sigma_{los}(\rh)$ may give a low estimate of the
satellite's true $\vm$. Not only do baryons change the dark matter
subhalos, as discussed in Section~\ref{sec:Results-TBTF}; but not all
subhalos are expected to host satellite galaxies, and the probability
for a low mass subhalo to host a satellite galaxy depends on its
$\vm$, and the typical stellar mass of a satellite galaxy also depends
on $\vm$. Hence, not all satellite halos have the same probability of
matching an observed satellite galaxy of a known stellar mass.

Indeed, the sample of satellite galaxies analysed by
\cite{Boylan-Kolchin-2012} contains nine of the twelve most luminous
known MW satellites, which are likely to be amongst the satellites
whose halos have the highest $\vm$ values. $\Lambda$CDM predicts
many more low $\vm$ halos than high $\vm$ halos. While
\citeauthor{Boylan-Kolchin-2012} point out that this does not affect
their analysis, it may amplify any potential bias caused by the
implicit assumption that all halos have an equal chance of hosting a
satellite of a given stellar mass.

In Fig.~\ref{fig:rotation-curves}, we examine the circular velocity
profiles of individual satellite halos and galaxies in {\sc Apostle},
and compare them to individual observed MW satellites. On each panel,
we show the circular velocity curves of satellites located within 300
kpc from the four central galaxies in two LG simulations at resolution
L1, and compare them to the circular velocities inferred from
$\sigma_{los}(\rh)$ for nine observed MW dwarf spheroidal galaxies.

On the top left panel of Fig.~\ref{fig:rotation-curves} (black
curves), we have selected satellite subhalos from the DMO simulation,
requiring that the circular velocity $\vc$ be within three times the
quoted observational uncertainty of $\sigma_{los}(\rh)$ for a given
satellite at its observed half-light radius, assuming the relation
between $\vc$ and $\sigma_{los}(\rh)$ of \cite{Wolf-2010}.

For the remaining three panels of Fig.~\ref{fig:rotation-curves}, we
have selected satellite subhalos from the corresponding hydrodynamic
simulations. In the top right (grey curves), we use the same selection
criterion as for the DMO case. For the bottom left panel (cyan
curves), we have not applied any velocity criterion, but require the
stellar mass in the simulation to be within $50\%$ of the observed
stellar mass of the individual satellite. Finally, in the bottom right
panel (dark blue curves), we combine the two previous criteria, and select
satellites from the hydrodynamic simulation whose stellar mass and
measured circular velocity are both compatible with those of the
observed satellite.

Due to the range in concentration of CDM halos, both the DMO and
hydrodynamic simulations allow a large range of $\vm$ values for most
observed satellites, in particular for those where the half-light
radius, $\rh$, is small compared to the radius where the circular
velocity is maximal, $\rmax$. It is also worth noting that when
satellites are selected purely by stellar mass, we find that many of
the simulated galaxies live in subhalos that are consistent with the
observed kinematics. As discussed in Section~\ref{sec:right}, a
notable exception is the Draco dSph, which appears to have an
unusually high halo mass for its stellar mass.

Combining the velocity and stellar mass criteria generally reduces the
range of compatible $\vm$ values for a particular satellite. We list
the most likely $\vm$ value for each of the nine observed satellites
according to our simulations in the two rightmost columns of Table
~\ref{table:vmax}, for both the DMO simulation using only the velocity
criterion, and for the hydrodynamic simulation, using both the
velocity and stellar mass constraints.

It should be noted that while the ranges quoted for the values of
\cite{Boylan-Kolchin-2012} denote the uncertainty of the most likely
value, for our own values we give the 16th and 84th percentiles of
$\vm$ for the compatible satellite halos (equivalent to $\pm 1 \sigma$
for a normal distribution). The latter can be much larger, and reflect
the diversity of halos in $\Lambda$CDM and in our simulations. It is
also worth pointing out again that any of the quoted $\vm$ estimates
from observed velocity dispersions are only valid in the $\Lambda$CDM
context and, for the values that also use the stellar mass, under the
additional assumptions made in our simulations.

\subsection{The right satellites in the right halos}\label{sec:right}
We have shown in Figs.~\ref{fig:smf} and ~\ref{FigVmaxFunc} that the
{\sc Apostle} simulations reproduce the number of satellites as a
function of stellar mass and of $\vm$, as inferred from the stellar
velocity dispersion and sizes. However, this success does not
necessarily imply that the simulations reproduce the stellar
mass--$\vm$ relation for individual satellites. Furthermore, the
stellar-to-total mass ratios of individual dwarf spheroidals show a
surprising amount of scatter: Fornax is roughly 100 times brighter
than Draco, but appears to inhabit a halo of similar mass.

In our simulations, the key to understanding this puzzling result lies
in the fact that, in the $\vm$ range below 30 km/s, where some halos
remain dark, those satellite halos that host galaxies similar to the
Milky Way dwarf spheroidals have typically experienced much more
severe tidal stripping, with a resulting reduction in mass or $\vm$
greater than expected for typical halos of the same mass
\citep{Sawala-2014b}. This phenomenon is illustrated in
Fig.~\ref{fig:mstar-mhalo}, which compares the stellar mass--$\vm$
relation of both central and satellite galaxies in our simulations
with data for the actual satellites of the Milky Way. While isolated
halos in our simulations fall below the stellar mass--$\vm$ relation
of observed satellites, the strong tidal stripping experienced by the
satellite galaxies and their halos brings them into good agreement
with the observed MW dwarf spheroidal data. The effect of stripping
can also account for the large scatter in stellar mass of dwarf
galaxies in a very narrow $\vm$ range. In this scenario, the high
stellar mass--$\vm$ ratio of satellites like Fornax and Sextans would
be explained in part by tidal stripping. By contrast, the low
stellar-to-halo mass ratio of Draco is more typical of isolated dwarf
galaxies in our simulations, suggesting that Draco has not yet
experienced strong tidal effects.

\begin{figure}
\begin{center}
  \resizebox{8.5cm}{!}{\includegraphics{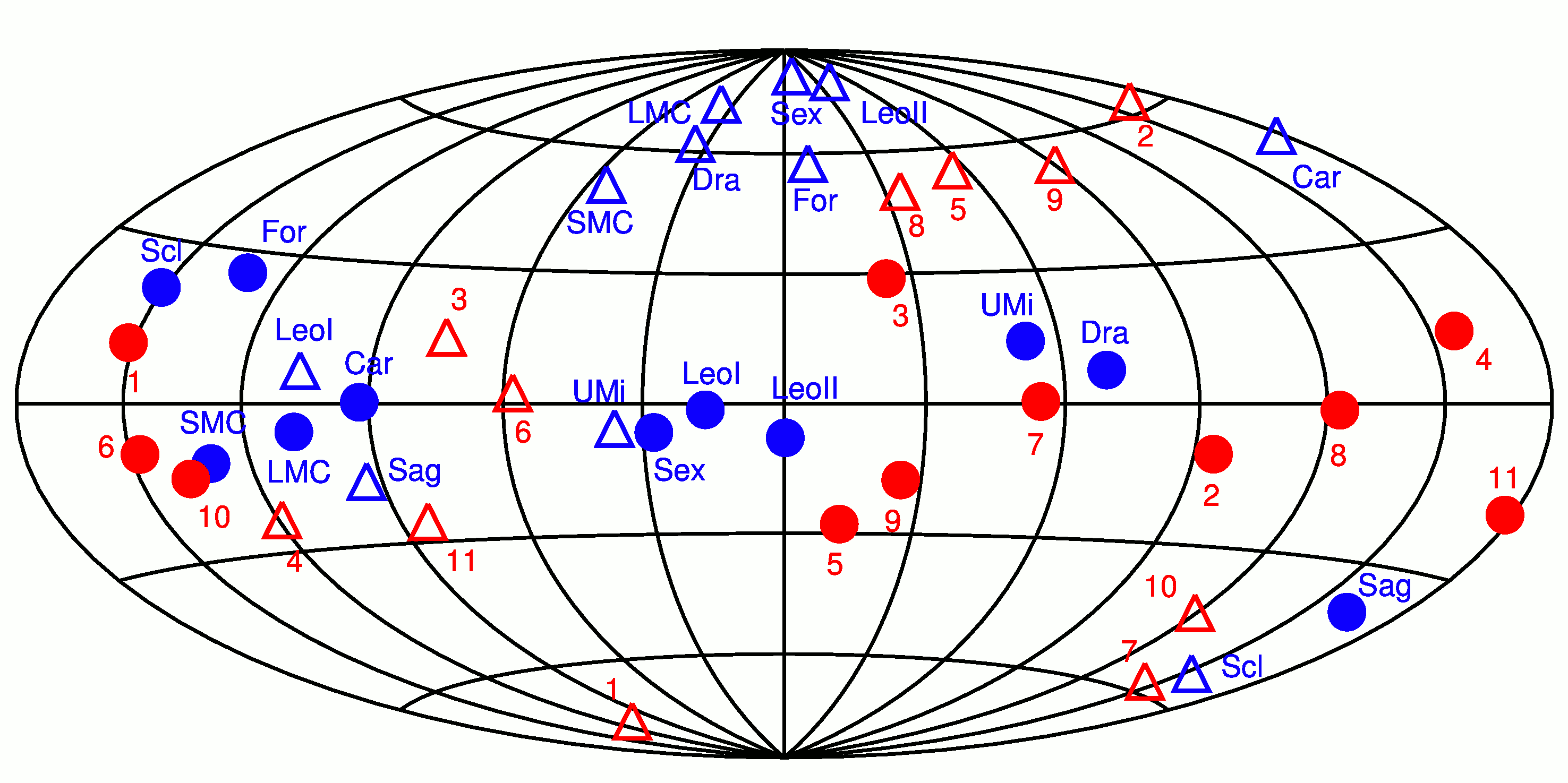}}
\caption{Angular distribution and kinematics of the eleven brightest satellites for a
  Milky-Way like system in our simulation (red), and for the observed
  Milky Way satellites (blue). Circles show the 
location on the sky, while triangles 
indicate the orientation of the corresponding angular momentum
vector. For both systems, the equator is chosen
  to align with the respective plane of satellites. The eleven brightest satellites in the simulated system are
distributed on a plane just as flat as those of the Milky Way and
several of them have a coherent rotation. Selecting the brightest satellites, systems as
  anisotropic as the MW's can be formed in $\Lambda$CDM, although
  they are not typical.
  \label{FigAnis}}
\end{center}
\end{figure}

In summary, not only do our hydrodynamic simulations reproduce both
the observed satellite stellar mass function and satellite circular
velocity function; but satellite galaxies of stellar masses comparable to
observed dwarf spheroidals also live in halos with compatible velocity
profiles. The reduction in subhalo mass due to baryonic effects, and
the strong stripping of halos that host the luminous satellites,
combine to give a satellite population that not only matches the Milky
Way and M31 satellite luminosity function, but also the total velocity
function of the observed satellite population.

\subsection{Unsurprisingly aligned}\label{sec:Results-Plane}
Finally, the anisotropy, first noticed by \cite{Lynden-Bell-1976}, and the
apparent orbital alignment of the 11 brightest, so-called
``classical'' MW satellites, have been regarded as highly improbable
in $\Lambda$CDM \citep[e.g.][]{Pawlowski-2013}. Fig.~\ref{FigAnis} compares
the observed angular distribution and orbital kinematics of the 11
brightest Milky Way satellites to the 11 brightest satellites around
one of our simulated LG galaxies at $z=0$, as identified in
Fig.~\ref{FigLGImage}. Both the simulated and observed satellite
populations show highly anisotropic distributions.

To characterise the anisotropy of each satellite system, we compute
the ratio of the minimal to maximal eigenvalues, $c$ and $a$, of the
reduced inertia tensor \citep[e.g.][]{Bailin-2005} defined by the 11
brightest satellites, ${I_{\alpha,\beta} =
  \sum_{i=1}^{11}{r_{i,\alpha}r_{i,\beta}} / r_i^2}$.  From a total of
24 MW or M31-like halos in our 12 LG simulations at resolution L2, we
find values of $\sqrt{c/a}$ in the range 0.34 -- 0.67, compared to
0.36 for the MW and 0.53 for M31.

Clearly, $\Lambda$CDM can produce satellite systems with a range of
anisotropies, and consistent with measurements for both the MW and M31.

However, considering only the 11 brightest satellites, the satellite
distribution of the MW appears to be more anisotropic than all but one
of our 24 systems, which still leaves open the question of its
statistical significance. It should be noted that, by the same
criterion, the satellite system of M31 is much less anisotropic than
that of the MW, and indeed quite typical of our simulated LG
galaxies. Given that the MW and M31 clearly formed within the same
cosmology with different satellite anisotropies, it may be premature
to consider the anisotropy of one a failure of the cosmological model.

The inclusion of baryonic physics in the {\sc Apostle} simulations
allow us to directly identify the brightest satellite galaxies out of
a much larger number of satellite halos. However, their spatial
anisotropy is clearly not caused by baryonic effects; it is intrinsic
to the assembly of $\Lambda$CDM satellite systems. As
\cite{Cautun-2015} have recently demonstrated, even when subsets of
observed satellites are carefully chosen to exhibit maximal anisotropy
\citep[e.g.][]{Ibata-2014}, rigorous statistical analysis shows that
the observed spatial and kinematic anisotropies are not inconsistent
with $\Lambda$CDM and that such apparently ``unusual'' systems are in
fact quite ubiquitous.

That satellite alignments are consistent with, and indeed expected to
exist in $\Lambda$CDM does not render them uninteresting. Each
individual satellite system can still contain information about the
assembly history of the halo and its relation to the large scale
structure. Across our simulations, the satellite plane most like the
MW's is aligned with a cosmic filament that envelopes the Local
Group. This supports the scenario proposed by \cite{Libeskind-2005},
whereby the accretion of satellites from the ``cosmic web'' imparts a
degree of coherence to the timing and direction of satellite
accretion.

\section{Summary}\label{sec:Summary}
{\sc Apostle} is a series of high-resolution zoom simulations of
the Local Group within the $\LCDM$ cosmology, performed both as dark
matter only, and including baryonic processes with the \eagle code. We
find that our simulations accurately reproduce the observed stellar
mass function of the Local Group, and also result in satellite
populations that are in good agreement with the observed relation
between dwarf galaxies and their dark matter halos.

We conclude that the ``problems'' often cited as challenges to
$\Lambda$CDM are resolved in simulations that reproduce the dynamical
constraints of the Local Group environment and include a realistic
galaxy formation model. Reionisation and supernova feedback allow
galaxy formation to proceed only in a small subset of dark matter
halos, eliminating the ``missing satellites'' problem. It is notable
that the very same galaxy formation model calibrated to reproduce the
galaxy population in cosmologically representative volumes naturally
produces a LG galaxy population in volumes consistent with the LG
kinematics in $\Lambda$CDM.

The loss of baryons due to reionisation and star formation feedback,
and from satellites through ram pressure stripping, affects the growth
of low-mass halos, leading to a reduction in their maximum circular
velocity compared to a DMO simulation. Combined with the effect of
tidal stripping, strongly enhanced for luminous, low mass halos, this
not only resolves the ``too-big-to-fail'' problem, but it also leads
to a satellite circular velocity function that matches observations,
at the same time as it matches the observed stellar mass
function. Furthermore, kinematics of individual simulated satellites
are consistent with the kinematics of observed satellites matched by
stellar mass. We also find that individual galaxies formed in our
simulations follow the observed relation between $\vm$ and stellar
mass, and use our DMO and hydrodynamic simulations to estimate the
range of likely $\vm$ values for nine dwarf spheroidals with measured
stellar mass and $\sigma_{los}(\rh)$ in the $\Lambda$CDM cosmology.

Finally, we find a diversity of satellite systems with spatial
anisotropy similar to those of the M31, and one system that is similar
to the ``plane of satellites'' around the MW. We conclude that the
observed anisotropies of these satellite systems do not falsify
$\Lambda$CDM, but may reflect their assembly histories within the
$\Lambda$CDM paradigm.

Another often-cited difficulty for $\Lambda$CDM is the inference of
constant density cores in dark matter halos
\citep[e.g.][]{Walker-2011}, whereas N-body simulations predict
cusps. While the observed kinematics of LG dwarf spheroidals are, in
fact, consistent with either cores or cusps \citep{Strigari-2014}, it
has also been argued that cores are required to solve the
too-big-to-fail problem \citep{Brooks-2013}. This is not the case: the
star formation and feedback model in our simulations and the effect of
tidal stripping give rise to a realistic LG galaxy population without
cores. Hence, we conclude that cores are not necessary to solve the
perceived small scale problems of $\Lambda$CDM.

Our simulations predict that the relation between stellar mass and
$\vm$ should differ between satellites and isolated dwarf galaxies, as
the observed satellites with high stellar mass--$\vm$ ratios live in
halos that experienced particularly strong tidal stripping. This
prediction is testable: alternative scenarios in which the dark matter
halos are modified independently of environment should give much more
similar relations between the two.

Our simulations are still limited by resolution and the shortcomings
of the subgrid physics model. Nevertheless, they result in galaxy
populations compatible with many of the observations of the LG galaxy
population and do not suffer from any of the problems often
interpreted as a failure of $\Lambda$CDM. Using a model already shown
to reproduce the galaxy population on much larger scales, they suggest
that the success and predictive power of the $\Lambda$CDM cosmology
extend far into the LG regime, once the effects of galaxy formation
and of the particular LG environment are taken into account.

\section*{Acknowledgements}
We are indebted to Dr. Lydia Heck who looks after the supercomputers
at the ICC. This work was supported by the Science and Technology
Facilities Council [grant number ST/F001166/1 and RF040218], the
European Research Council under the European Union's Seventh Framework
Programme (FP7/2007-2013) / ERC Grant agreement 278594
'GasAroundGalaxies', the National Science Foundation under Grant
No. PHYS-1066293, the Interuniversity Attraction Poles Programme of
the Belgian Science Policy Office [AP P7/08 CHARM]. T. S. acknowledges
the Marie-Curie ITN CosmoComp and the support of the Academy of
Finland grant 1274931. C.~S.~F. acknowledges ERC Advanced Grant 267291
'COSMIWAY' and S.~W. acknowledges ERC Advanced Grant 246797
'GALFORMOD'. R.~A.~C. is a Royal Society University Research
Fellow. This work used the DiRAC Data Centric system at Durham
University, operated by the Institute for Computational Cosmology on
behalf of the STFC DiRAC HPC Facility (www.dirac.ac.uk), and resources
provided by WestGrid (www.westgrid.ca) and Compute Canada / Calcul
Canada (www.computecanada.ca). The DiRAC system is funded by BIS
National E-infrastructure capital grant ST/K00042X/1, STFC capital
grant ST/H008519/1, STFC DiRAC Operations grant ST/K003267/1, and
Durham University. DiRAC is part of the National E-Infrastructure.

\bibliographystyle{mn2e} \bibliography{paper}

\appendix

\section{Numerical Convergence}

\begin{figure*}
\begin{center}
\includegraphics[width=8.3cm, trim={1.cm 0 1.cm 1.5cm},clip]{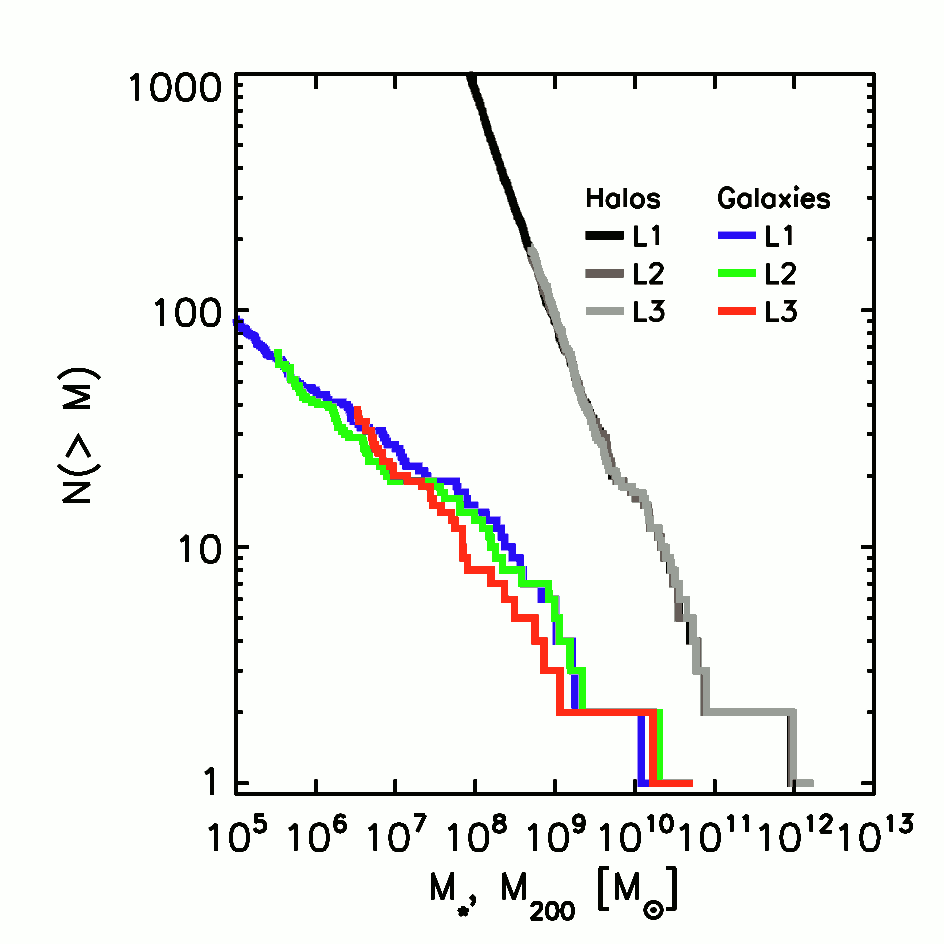}
\includegraphics[width=8.3cm, trim={1.cm 0 1.cm 1.5cm},clip]{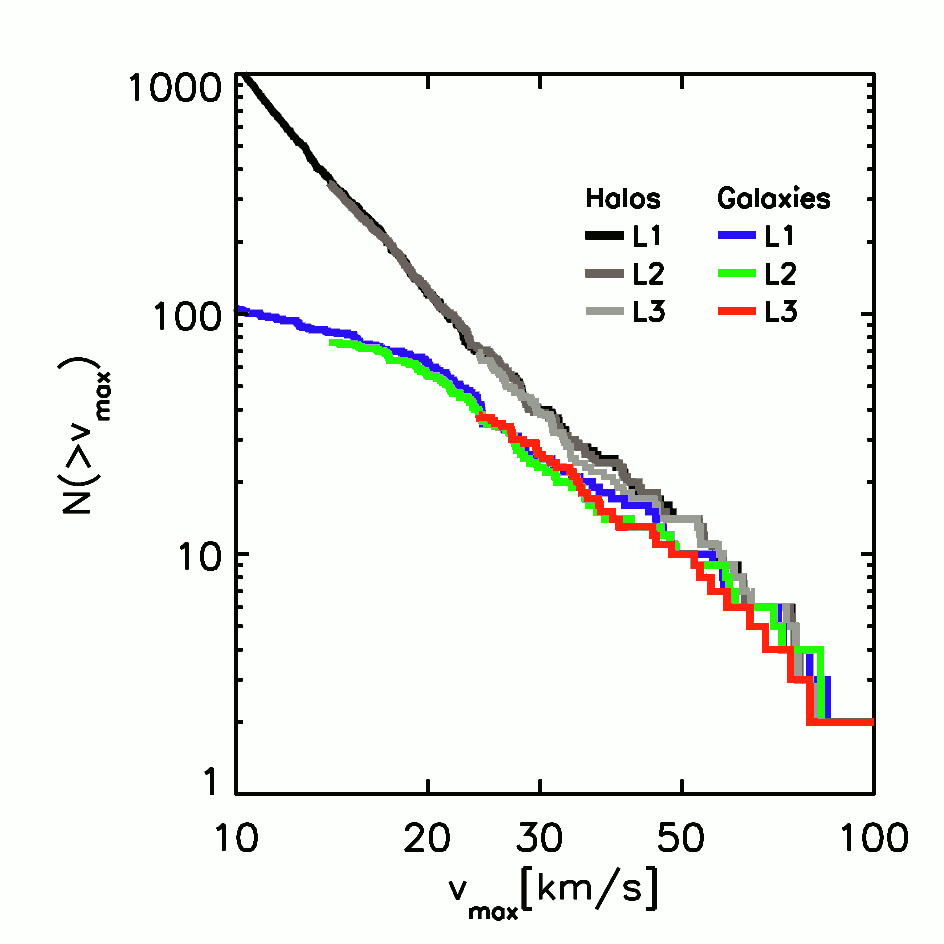} 
\caption{Stellar and halo mass functions (left panel) and maximum
  circular velocity functions (right panel) within 2 Mpc from the
  barycentre of one of the {\sc Apostle} volumes at three different
  resolutions, for galaxies in the hydrodynamic simulations, and halos
  in the corresponding DMO simulations. The stellar and halo mass
  functions, and the velocity functions of galaxies and halos are well
  converged at resolution L2.} \label{fig:convergence}
\end{center}
\end{figure*}

As discussed in Section~\ref{sec:Methods-ICs}, we have performed our
simulations at 3 resolution levels, varying by a factor of $\sim144$
in particle mass. Fig.~\ref{fig:convergence} illustrates the numerical
convergence of the mass functions (left panel) and velocity function
(right panel) for one {\sc Apostle} volume at the three resolution
levels. For both the halo mass function and halo velocity function,
results are shown from the DMO simulation. Here, convergence is
excellent. For the galaxy mass function, and the velocity function of
the halos containing galaxies (coloured curves in both figures),
results are shown from the hydrodynamic simulation.

Convergence of the cumulative stellar mass function remains very
good. By comparison to Fig.~\ref{fig:smf}, which includes results from
all twelve {\sc Apostle} volumes at resolution L2, it is clear that
the difference in the stellar mass function due to numerical
resolution is much smaller than the variation between the different
{\sc Apostle} volumes.

The velocity function of galaxies, which falls below the corresponding
total velocity function of halos in the DMO simulation due the effect
of baryons on halos at $\vm \sim 60$ km/s, and due to the appearance
of ``dark'' halos at t $\vm \sim 30$ km/s, is equally well
converged. In particular, both the average reduction in $\vm$ of
subhalos, and the fraction of dark halos, are independent of
resolution.

\label{lastpage}

\end{document}